\documentclass[aps,jor,groupedaddress]{revtex4-2}
\usepackage[margin=1in]{geometry}
\usepackage{amsmath,accents}
\usepackage[labelfont={small}, font={normal}]{subfig} 
\usepackage{graphicx}
\usepackage{epstopdf}
\usepackage{amssymb}
\usepackage{enumitem}
\usepackage[utf8]{inputenc}
\usepackage[dvipsnames]{xcolor}

\begin{document}
\title{Stress analysis of dilute particle suspensions in non-Newtonian fluids with efficient evaluation in the weakly non-Newtonian limit}
\author{Arjun Sharma}
\affiliation{Sibley School of Mechanical and Aerospace Engineering, Cornell University, Ithaca, NY, 14853, USA}
\author{Donald L. Koch\footnote{Corresponding author: {dlk15@cornell.edu}}}
\affiliation{Robert Frederick Smith School of Chemical and Biomolecular Engineering, Cornell University, Ithaca, NY, 14853, USA}	
\begin{abstract}
{We present a semi-analytical framework to compute the suspension stress in dilute particle-laden non-Newtonian fluids, separating Newtonian and non-Newtonian contributions. The ensemble-averaged stress includes both the particle-induced non-Newtonian stress (PINNS) and an interaction stresslet arising from surface tractions due to the non-Newtonian stress and its induced Newtonian flow. Using a generalized reciprocal theorem, we express this interaction stresslet entirely in terms of the non-Newtonian stress, for a general constitutive model. For weakly non-Newtonian fluids, a regular perturbation expansion combined with the method of characteristics yields all leading-order stress contributions from the Newtonian velocity field alone, avoiding the need to solve coupled partial differential equations. This generalizes the method of Koch et al. [Phys. Rev. Fluids 1, 013301 (2016)] beyond polymeric fluids to any weakly non-Newtonian medium driven by velocity and its gradients. We apply the method to two systems: (i) spheres suspended in a fluid of smaller spheroids, where the interaction stress becomes negative for sufficiently anisotropic shapes due to orientation misalignment of the spheroids; and (ii) suspensions in weakly anisotropic nematic liquid crystals. In the latter, assuming a uniform director field fixed by an external field, PINNS vanishes while interaction stresslets remain, either opposing or enhancing background anisotropic stress. These results demonstrate the utility of our framework in capturing first-order particle–microstructure interactions across a broad class of non-Newtonian fluids.}
\end{abstract}
\maketitle
	
\section{Introduction}\label{sec:Introduction}
Suspensions of particles in non-Newtonian fluids arise across a wide range of systems, including everyday consumer products (e.g., toothpaste, shampoos), industrial processes (e.g., hydraulic fracturing, fiber spinning, and low-resistance film manufacturing), geophysical flows (e.g., landslides, mudflows, snow avalanches, and volcanic lava), and biological environments (e.g., blood and mucus)~\cite{breitenbach2002melt,huang2003review,nakajima1994advanced,chae2008making,mutiso2013integrating,yin2010inkjet,barbati2016complex}. The suspending media in these applications are often viscoelastic or elastoviscoplastic, with rheological properties such as relaxation time and yield stress leading to a broad range of complex, mechanistically distinct behaviors, especially when coupling between suspended particles and the fluid microstructure (e.g., polymer chains or mesogens) is considered. Beyond naturally occurring systems, synthetic non-Newtonian fluids with tunable rheological properties have also been engineered for targeted applications~\cite{ewoldt2022designing}. In this work, we develop a unified formulation for computing the rheology of dilute suspensions in a broad class of non-Newtonian fluids, generalizing Batchelor’s classical framework for Newtonian suspensions~\cite{batchelor1970stress}. Specifically, we express the non-Newtonian contributions to the suspension stress explicitly in terms of the base fluid’s non-Newtonian stress tensor, {for any fluid whose stress can be decomposed into a Newtonian part and a non-Newtonian part. Importantly, the formulation itself does not assume weak non-Newtonian behavior. However, because the non-Newtonian contribution to the suspension stress depends only on the base-fluid non-Newtonian stress, in the weakly non-Newtonian limit a regular perturbation expansion allows us to circumvent the need to evaluate intermediate (first-order) disturbance fields such as the  velocity and pressure.} {Hence, this formulation enables computationally efficient estimation of suspension stresses and offers a tool for early-stage rheological characterization of novel materials.} We present two illustrative examples using this methodology.

In dilute suspensions, where the particle volume fraction $\phi$ is small and interparticle interactions are negligible, ensemble averaging techniques~\cite{hinch1977averaged,koch2006stress,koch2016stress,sharma2023steady} allow the suspension stress to be determined from the flow of the base fluid around a single particle. For many classes of non-Newtonian fluids, such as viscoelastic~\cite{larson2013constitutive}, elastoviscoplastic~\cite{saramito2009new}, and liquid crystalline materials~\cite{de1993physics}, the fluid stress can be decomposed into Newtonian and non-Newtonian components. Our general framework builds on ensemble averaging methods previously developed for second-order~\cite{koch2006stress} and {polymeric~\cite{koch2016stress,sharma2023steady,sharma2025extensional}} fluids. The interaction of the particle with the non-Newtonian component introduces two key contributions to the suspension stress: (i) the particle-induced non-Newtonian stress (PINNS), which arises from the local distortion of the microstructure (e.g., polymer configuration or nematic alignment), and (ii) the interaction stresslet, which reflects the additional surface traction exerted by the fluid on the particle due to non-Newtonian effects. 

PINNS represents the non-Newtonian stress generated by the disturbance of the flow field around the particle. In viscoelastic fluids, it arises due to distortion of polymer configurations relative to those in the undisturbed flow; in liquid crystals, it reflects distortions of the nematic director field. The particle stresslet, which captures the integrated fluid traction on the particle surface, exists even in Newtonian suspensions. In non-Newtonian fluids, however, an additional component, referred to as the interaction stresslet, emerges due to coupling with the non-Newtonian stress. The total Newtonian stress in a non-Newtonian fluid can be viewed as the sum of a Newtonian contribution (i.e., what the stress would be in a Newtonian fluid) and an additional term that is induced by the divergence in the non-Newtonian stress. The interaction stresslet therefore includes contributions from both the non-Newtonian stress and this induced Newtonian stress. Although in general the microstructure (e.g., polymer melt) may not be explicitly separable from the Newtonian solvent, the total stress can still be expressed as the sum of Newtonian and non-Newtonian components, enabling this decomposition to be applied broadly.

We introduce a novel decomposition of the particle stresslet into a volumetric contribution and an “undisturbed stresslet” term that depends only on the particle geometry and the far-field or imposed flow field. The undisturbed stresslet corresponds to the stress exerted on a body of identical shape in an undisturbed flow and is given by the product of the symmetric part of the undisturbed stress and the particle volume. When used in combination with a generalized reciprocal theorem, this decomposition yields an expression for the interaction stresslet that depends only on the non-Newtonian stress field, making it both physically interpretable and computationally tractable.

{Two distinct features of our framework are central contributions of this work: (i) a careful decomposition of the ensemble-averaged non-Newtonian stress, applicable to a wide class of non-Newtonian fluids, into the particle-induced non-Newtonian stress (PINNS) and the undisturbed non-Newtonian stress; and (ii) an expression for the particle stresslet as an explicit functional (integral operator) of the non-Newtonian stress field alone. Our detailed treatment of how the ensemble-averaged suspension stress relates to a measurable volume average is especially important for non-Newtonian fluids, where the stress depends nonlinearly on the flow. Previous analyses of viscoelastic suspensions \citep{greco2007rheology,housiadas2009rheology,jain2019extensional} have directly taken the volume average of the non-Newtonian stress in place of the ensemble average; we show that this substitution can be problematic. The particle stresslet arises from (i) the non-Newtonian stress acting on the particle surface and (ii) the perturbation to the surface Newtonian stress induced by the non-Newtonian stress. The alternative decomposition described above circumvents the need to evaluate the latter explicitly by expressing it directly in terms of the non-Newtonian stress, yielding substantial computational savings in the weakly non-Newtonian fluids.}

In general, the constitutive equations governing non-Newtonian stress are coupled with the momentum balance equations, requiring fully resolved numerical simulations that are often computationally expensive and prone to instabilities due to steep gradients (e.g., in polymer stretch). These challenges restrict the parameter regimes that can be explored and hinder the investigation of novel or hypothetical material systems. To address these limitations, we develop a semi-analytical method tailored to weakly non-Newtonian fluids, where the non-Newtonian stress is formally $\mathcal{O}(\epsilon)$ (with $\epsilon$ being the weak non-Newtonian parameter for example polymer concentration) smaller than the Newtonian stress. Such regimes occur, for instance, in viscoelastic fluids with low polymer concentrations or in nematic liquid crystals near the isotropic–nematic transition, where anisotropic stresses are weak~\cite{chandrasekar2023micro}. Importantly, this weakly non-Newtonian assumption does not constrain the relaxation time or other intrinsic properties of the microstructure. The method thus preserves physical richness while dramatically reducing numerical complexity.

In this perturbative regime, the leading-order Newtonian velocity field governs the first-order change in the non-Newtonian stress, and all $\mathcal{O}(\phi \epsilon)$ corrections to the suspension stress can be evaluated without solving the coupled flow equations. Because our stresslet formulation depends only on the non-Newtonian stress, the remainder of the Newtonian stress, ordinarily determined by solving a partial differential equation enforcing momentum balance, need not be explicitly computed. For many constitutive models, the non-Newtonian stress depends on the velocity and its gradients, and its governing equation takes the form of a PDE with convective derivatives. We recast such equations as ordinary differential equations (ODEs) along streamlines of the Newtonian flow using the method of characteristics. This reduces the computational cost to that of solving a set of ODEs, enabling high-resolution computation of the non-Newtonian stress across a broad non-Newtonian parameter space.

To demonstrate the method, we examine two model systems: (i) a spheroidal fluid, defined as a dilute suspension of small, non-Brownian spheroids in a Newtonian solvent, and (ii) a nematic liquid crystal with weak anisotropy. In both cases, the suspended spheres are significantly larger than the microstructural elements, allowing us to isolate particle–microstructure interactions. These examples are chosen for the tractability of their constitutive equations and the physical richness of their microstructure–flow coupling. The spheroidal fluid model, for example, has been used previously to study drag reduction in turbulent flows with fiber additives~\cite{paschkewitz2004numerical}. {In the case of liquid crystals, prior work has highlighted the potential of particle–LC suspensions to elucidate fundamental problems in soft matter physics and guide applications in reconfigurable materials~\cite{stark2001physics,smalyukh2018liquid}. For suspensions in weakly anisotropic nematic fluids, the entire calculation is carried out analytically, and we identify novel contributions to the suspension stress that result from coupling between the particle and the anisotropic stress field.} {Thus, this semi-analytical method can be used to design new non-Newtonian fluids or to rapidly explore existing systems in previously unexplored flow regimes, complementing and guiding experimental studies aimed at this effort \citep{ewoldt2022designing}.}

The rest of the paper is organized as follows. In section~\ref{sec:FormulationAllNonNewtonianFluid}, we present the general formulation for calculating the ensemble-averaged stress in dilute suspensions in non-Newtonian fluids and introduce the novel stresslet decomposition. Section~\ref{sec:SuspensionStressWeakNonNewtonian} details simplifications that arise in the weakly non-Newtonian regime and the resulting semi-analytical method. In section~\ref{sec:examples}, we apply the framework to the spheroidal and liquid crystal fluids described above. Finally, section~\ref{sec:Conclusions} summarizes the key findings.

\section{Ensemble-Averaged Stress in a Suspension within a Non-Newtonian Fluid}\label{sec:FormulationAllNonNewtonianFluid}
We consider the flow of an inertia-less, incompressible, non-Newtonian fluid containing a dilute suspension of particles. The conservation of mass and momentum is enforced through divergence-free conditions on the velocity field, $\mathbf{u}$, and the total stress tensor, $\boldsymbol{\sigma}$, throughout the suspension:
\begin{eqnarray}
	\nabla\cdot \mathbf{u}=0,\hspace{0.2in}
	\nabla\cdot \boldsymbol{\sigma}=0.\label{eq:momentum}
\end{eqnarray}
The stress at any point within the suspension is given by:
\begin{equation}\label{eq:constitutive3}
	\boldsymbol{\sigma}=\boldsymbol{\tau}+\boldsymbol{\Pi}+\boldsymbol{\sigma}^\text{E}=-p\boldsymbol{\delta}+2\mathbf{e}+\boldsymbol{\Pi}+\boldsymbol{\sigma}^\text{E},
\end{equation}
where $\boldsymbol{\sigma}^{\text{E}}$ denotes the extra stress within a particle, which vanishes in the fluid region. The fluid stress consists of a Newtonian component, $\boldsymbol{\tau}$, and a non-Newtonian contribution, $\boldsymbol{\Pi}$. In the Newtonian part, $p$ is the hydrodynamic pressure, $\boldsymbol{\delta}$ is the identity tensor, and $\mathbf{e}=(\nabla \mathbf{u}+\nabla \mathbf{u}^{\text{T}})/2$ is the rate-of-strain tensor.  All quantities are nondimensionalized using the particle length scale (specifically, the sphere radius in Section~\ref{sec:examples}) and the inverse of a characteristic imposed deformation rate (e.g., shear or extension, {$\dot{\epsilon}$}) as the time scale, {and the solvent viscosity $\eta$ to scale the fluid stress by $\eta\dot{\epsilon}$, so that the Newtonian viscous stress appears as $2\mathbf{e}$}. Equation~\eqref{eq:momentum} is subject to no-slip and no-penetration conditions on the particle surface and specified flow conditions at the suspension boundaries.

The isotropic component of the stress, $\operatorname{tr}(\boldsymbol{\sigma})/3$, can be absorbed into the pressure, leading to the deviatoric stress:
\begin{equation}\label{eq:DeviatoricStress}
	\hat{\boldsymbol{\sigma}}
	= \boldsymbol{\sigma} - \frac{\operatorname{tr}(\boldsymbol{\sigma})}{3}\,\boldsymbol{\delta}
	= 2\mathbf{e} + \hat{\boldsymbol{\Pi}} + \hat{\boldsymbol{\sigma}}^{\text{E}}.
\end{equation}
To evaluate the rheology of the suspension, we apply ensemble averaging \cite{koch2006stress,koch2016stress,hinch1977averaged}. For completeness, we follow the derivation in \cite{koch2016stress}. {As noted by \cite{batchelor1970stress}, the suspension stress observed in an experiment is an ensemble average over all possible realizations of particle configurations. However, this average is not directly measurable and is typically replaced by a volume average via the ergodic hypothesis. In non-Newtonian fluids, where the fluid stress (specifically $\boldsymbol{\Pi}$) may depend nonlinearly on the velocity, the replacement must be introduced at the correct stage. We therefore begin with the ensemble-average definition and then describe the proper way to replace the ensemble average with a volume average. This yields the familiar stresslet (present also for Newtonian fluids) as one of the components of particle-induced stresses in the suspension. The other component specific to non-Newtonian fluids is the particle-induced non-Newtonian stress arising from the cumulative effect of the disturbance of $\boldsymbol{\Pi}$ by the particle.} The ensemble average (denoted $\langle \cdot \rangle$) of the deviatoric stress is
\begin{equation}\label{eq:SuspensionStress}
	\langle \hat{\boldsymbol{\sigma}} \rangle
	= 2\langle \mathbf{e} \rangle + \langle \hat{\boldsymbol{\Pi}} \rangle + n\,\hat{\mathbf{S}}(\boldsymbol{\sigma}),
\end{equation}
where $n$ is the particle number density, and the particle stresslet $\hat{\mathbf{S}}$ is defined by
\begin{equation}\label{eq:Stresslet}
	\langle \hat{\boldsymbol{\sigma}}^{\text{E}} \rangle
	= n\,\hat{\mathbf{S}}(\boldsymbol{\sigma})
	= \int_{\mathbf{r}_p} \mathrm{d}\mathbf{r}_1\;
	\langle \hat{\boldsymbol{\sigma}}^{\text{E}} \rangle_1(\mathbf{r}\,|\,\mathbf{r}_1)\,P(\mathbf{r}_1).
\end{equation}
{The ensemble average of $\hat{\boldsymbol{\sigma}}^{\text{E}}$ involves only a volume integral over the particle region as by definition this extra particle stress is zero within the fluid.}
{Equation \eqref{eq:constitutive3} is defined at an arbitrary location within the suspension therefore while replacing the ensemble average with a measurable quantity entire volume of the suspension (fluid+particle) is used. The above integral is performed over a particle volume bounded by the surface $\mathbf{r}_p$ as the extra particle stress $\hat{\boldsymbol{\sigma}}^\text{E}=0$ within the fluid domain.} {Here,} $P(\mathbf{r}_1)$ is the probability density of finding a particle at position $\mathbf{r}_1$. For a quantity $A$, the conditional ensemble average $\langle A \rangle_1(\mathbf{r} | \mathbf{r}_1)$, with one particle positioned at $\mathbf{r}_1$, is defined as:
\begin{equation}
	\langle A\rangle_1(\mathbf{r}|\mathbf{r}_1)=\int\text{d}\mathbf{r}_2\dots \mathbf{r}_N P(\mathbf{r}_2\dots \mathbf{r}_N|\mathbf{r}_1)A,
\end{equation}
where $P(\mathbf{r}_2 \dots \mathbf{r}_N \mid \mathbf{r}_1)$ is the conditional probability density in a suspension of $N$ particles.

In the dilute limit, where interparticle interactions are negligible, the conditional average reduces to the stress field around an isolated particle \cite{koch2016stress}, allowing us to drop the conditional notation in \eqref{eq:Stresslet}. Applying the divergence theorem to the volume integral of $\boldsymbol{\sigma}$ around an isolated particle {and noting that for $\nabla\cdot\boldsymbol{\sigma}=0$, $\nabla\cdot(\boldsymbol{\sigma}\mathbf{r})=\boldsymbol{\sigma}$} gives the stresslet as a surface integral \cite{batchelor1970stress}:
\begin{equation}
	\hat{\mathbf{S}}(\boldsymbol{\sigma})
	= \int_{\text{particle surface}} \mathrm{d}A\,\bigg[
	\frac{1}{2}\big(\,\mathbf{r}\mathbf{n}\cdot\boldsymbol{\sigma}
	+ \mathbf{n}\cdot\boldsymbol{\sigma}\,\mathbf{r}\,\big)
	- \frac{1}{3}\,\boldsymbol{\delta}\,\big(\mathbf{n}\cdot\boldsymbol{\sigma}\cdot\mathbf{r}\big)
	\bigg],
	\label{eq:Stresslet1}
\end{equation}
where $\mathbf{n}$ is the unit normal on the particle surface pointing into the fluid. This form emphasizes that the particle phase does not require a constitutive relation, as only surface stresses contribute.

Special care is required when approximating the ensemble average of the non-Newtonian stress, $\langle \boldsymbol{\Pi} \rangle$, by volume averaging. A direct volume average can be inaccurate \cite{koch2016stress}: for polymeric fluids around a sphere, $\boldsymbol{\Pi}$ decays as $1/r^3$, leading to a logarithmic divergence in the volume integral. Here we outline a more careful treatment for evaluating $\langle \boldsymbol{\Pi} \rangle$ for a broader class of non-Newtonian fluids {(not only weakly non-Newtonian)}. {While this scaling argument offers a brief overview of how the far-field disturbance produces a logarithmic contamination of the suspension stress, an explicit calculation of the extensional rheology of dilute suspensions of spheres in polymeric liquids was recently provided by \cite{sharma2025extensional}. There, the volume average of the numerically evaluated polymeric stress failed to converge as the computational domain size increased, even for domains as large as 800 particle radii, whereas the more appropriate treatment, consistent with the general approach outlined below, achieved convergence for domains on the order of 200 particle radii.}

We decompose $\boldsymbol{\Pi}$ into undisturbed, linear, and nonlinear components:
\begin{equation}
	\boldsymbol{\Pi}=\boldsymbol{\Pi}^{U}+\boldsymbol{\Pi}^{L}+\boldsymbol{\Pi}^{NL}.
\end{equation}
{Here, $\boldsymbol{\Pi}^{U}$ is the non-Newtonian stress undisturbed by the test particle at $\mathbf{r}_1$. For dilute particle suspensions (with large inter-particle distances), when evaluating the rheology up to the leading order in particle volume fraction, it is equivalent to the stress in the absence of particles and may be termed as the undisturbed non-Newtonian stress. $\boldsymbol{\Pi}^{L}$ is the linearized non-Newtonian stress,}
where the linearization is taken with respect to the disturbance from the imposed or ensemble-averaged flow:
\begin{equation}\label{eq:UspletEnsemble}
	\mathbf{u}=\langle \mathbf{u} \rangle + \mathbf{u}' ,
\end{equation}
{and $\boldsymbol{\Pi}^{NL}$ is the remaining nonlinear stress. In the above equation, $\langle \mathbf{u} \rangle$ is the velocity field in the absence of the test particle and $\mathbf{u}'$ is the perturbation about this flow. Similar to that for $\boldsymbol{\Pi}$ mentioned above, for dilute particle suspensions, $\langle \mathbf{u} \rangle$ is the undisturbed or the imposed fluid velocity field. Otherwise, it is the ensemble average velocity, unconditional on the test particle.}
The constitutive equation for the non-Newtonian fluid, which depends on $\boldsymbol{\Pi}$ and $\mathbf{u}$, can be written generically as
\begin{equation}\label{eq:GeneralConstitutive}
	\mathbf{H}(\boldsymbol{\Pi},\mathbf{u})=0,
\end{equation}
where $\mathbf{H}$ may be a tensor- or vector-valued operator, as in models such as Oldroyd-B, FENE-P, Giesekus, Ericksen–Leslie. {The linearized stress $\boldsymbol{\Pi}^{L}$ is obtained by linearizing the constitutive equation with respect to the velocity field, i.e., by ignoring products of $\mathbf{u}'$ with itself. Therefore,} linearizing yields
\begin{equation}\label{eq:GenLinConst2}
	\tilde{\mathbf{H}}_{\boldsymbol{\Pi}^{L}}(\langle \mathbf{u} \rangle)\circ \boldsymbol{\Pi}^{L}
	+ \tilde{\mathbf{H}}_{\mathbf{u}'}(\boldsymbol{\Pi}^{U})\circ \mathbf{u}' = 0,
\end{equation}
where $\tilde{\mathbf{H}}_{\boldsymbol{\Pi}^{L}}$ and $\tilde{\mathbf{H}}_{\mathbf{u}'}$ are linear operators. Ensemble averaging gives
\begin{equation}\label{eq:GenLinConst3}
	\tilde{\mathbf{H}}_{\boldsymbol{\Pi}^{L}}(\langle \mathbf{u} \rangle)\circ \langle \boldsymbol{\Pi}^{L} \rangle
	+ \tilde{\mathbf{H}}_{\mathbf{u}'}(\boldsymbol{\Pi}^{U})\circ \langle \mathbf{u}' \rangle = 0.
\end{equation}
Since $\langle \mathbf{u}' \rangle = 0$ by definition of the ensemble average, it follows that
\begin{equation}\label{eq:GenLinConst4}
	\tilde{\mathbf{H}}_{\boldsymbol{\Pi}^{L}}(\langle \mathbf{u} \rangle)\circ \langle \boldsymbol{\Pi}^{L} \rangle = 0.
\end{equation}
{We consider the boundaries to be an artificial outer boundary used to truncate an otherwise unbounded suspension. Here, the velocity disturbance due to the test particle decays to zero as the distance from the particle tends to infinity. Consequently the non-Newtonian stress, $\boldsymbol{\Pi}\rightarrow\boldsymbol{\Pi}^{U}$, so that $\boldsymbol{\Pi}^{L}\rightarrow0$. In the numerical implementation we therefore impose $\boldsymbol{\Pi}^{L}=0$ at an outer boundary placed sufficiently far from the test particle. Thus, equation \eqref{eq:GenLinConst4} for $\langle \boldsymbol{\Pi}^{L} \rangle$, together with homogeneous boundary condition at the far-field boundary implies $\langle \boldsymbol{\Pi}^{L} \rangle = 0$.} Hence,
\begin{equation}
	\langle \boldsymbol{\Pi} \rangle = \boldsymbol{\Pi}^{U} + \langle \boldsymbol{\Pi}^{NL} \rangle .
\end{equation}

In analogy with the stresslet derivation, the nonlinear part of the non-Newtonian stress in a dilute suspension is approximated by a volume integral for an isolated particle:
\begin{equation}\label{eq:PINNS}
	V_{\text{particle}}\,\mathrm{PINNS}(\boldsymbol{\Pi},\boldsymbol{\Pi}^{U},\boldsymbol{\Pi}^{L})
	\;=\; \langle \boldsymbol{\Pi}^{NL} \rangle
	\;=\; n \int_{V_f + V_p} \boldsymbol{\Pi}^{NL}\, dV,
\end{equation}
where $V_f$ and $V_p$ denote the fluid and particle domains, and $V_{\text{particle}}$ is the particle volume.

{We may also consider a similar procedure for evaluating the ensemble average of the deviatoric Newtonian stress, i.e., the strain rate $\mathbf{e}$ under our nondimensionalization. Since the strain rate is linear in the velocity, $\mathbf{e}=\mathbf{e}^{U}+\mathbf{e}'$ with $\mathbf{e}'=(\nabla \mathbf{u}'+\nabla \mathbf{u}'^{\text{T}})/2$ the disturbance strain rate (already linear in the particle-induced disturbance), and $\mathbf{e}^{U}$ the imposed strain rate (the particle-free value), the nonlinear component of the Newtonian stress is zero. Because $\langle \mathbf{e}' \rangle = 0$ follows directly from $\langle \mathbf{u}' \rangle = 0$, we may express the ensemble average of the Newtonian stress as the undisturbed/imposed value: $\langle \mathbf{e} \rangle = \mathbf{e}^{U}$. It has been recognized that, due to the nonlinear dependence of $\boldsymbol{\Pi}$ on the velocity field, additional terms for viscoelastic liquids beyond the undisturbed stress $\boldsymbol{\Pi}^{U}$ will appear \citep{greco2007rheology,housiadas2009rheology,jain2019extensional}. However, a straightforward substitution of the ensemble average of $\boldsymbol{\Pi}$ by the volume average of the disturbance non-Newtonian stress $\boldsymbol{\Pi}-\boldsymbol{\Pi}^{U}$ is inappropriate because the linearized stress has an ensemble average of zero yet yields a logarithmically divergent volume integral. In numerical studies this manifests as a lack of convergence with respect to domain size, as shown in our recent work on the extensional rheology of polymeric liquids, where a naive volume average was contrasted with the volume average of the nonlinear stress \cite{sharma2025extensional}. Non-linearity in $\boldsymbol{\Pi}$ may also arise due to additional microstructural mechanisms such as the velocity-director coupling in nematic liquid crystals.}
	

Combining all contributions, the ensemble-averaged deviatoric stress is:
\begin{equation}\label{eq:SuspensionStress2}
	\langle\hat{\boldsymbol{\sigma}}\rangle = 2\langle\mathbf{e}\rangle + \hat{\boldsymbol{\Pi}}^U + n\int_{V_f+V_p}dV\text{  } \boldsymbol{\Pi}^{NL} + n\hat{\text{\textbf{S}}}(\boldsymbol{\sigma}).
\end{equation}
Here, the first term is twice the mean strain rate; the second, $\hat{\boldsymbol{\Pi}}^U$, is the undisturbed non-Newtonian stress in the absence of particles; the third term, termed ``particle-induced non-Newtonian stress" (PINNS), accounts for nonlinear deviations caused by particles; and the final term is the particle number density times the stresslet from surface traction. As shown in \eqref{eq:Stresslet1}, the stresslet is a linear function of $\boldsymbol{\sigma}$ and can be further decomposed as discussed next. Following this decomposition and using a generalized reciprocal theorem, the entire non-Newtonian contribution to the stresslet can be expressed as an explicit function of $\hat{\boldsymbol{\Pi}}$ and $\hat{\boldsymbol{\Pi}}^U$.

\subsection{Decomposition of the particle stresslet in non-Newtonian fluids}\label{sec:PhysicalDefinition}
As previously discussed, in a non-Newtonian fluid, the fluid stress on the particle surface comprises both Newtonian and non-Newtonian components. Accordingly, the stresslet can be expressed as:
\begin{equation}\label{eq:StressletNewtNonNewt}
	\hat{\text{\textbf{S}}}(\boldsymbol{\sigma}) = \hat{\text{\textbf{S}}}(\boldsymbol{\tau} + \boldsymbol{\Pi}) = \hat{\text{\textbf{S}}}(\boldsymbol{\tau}) + \hat{\text{\textbf{S}}}(\boldsymbol{\Pi}),
\end{equation}
where $\hat{\text{\textbf{S}}}(\boldsymbol{\tau})$ and $\hat{\text{\textbf{S}}}(\boldsymbol{\Pi})$ denote the stresslets arising from the Newtonian and non-Newtonian components, respectively.

To satisfy momentum conservation, the Newtonian stress $\boldsymbol{\tau}$ is perturbed by the presence of $\boldsymbol{\Pi}$. Thus, we write:
\begin{equation}\label{eq:StressletNewtintoNewt+NonNewtPert}
	\boldsymbol{\tau} = \boldsymbol{\tau}^N + \boldsymbol{\tau}^{NN},
\end{equation}
where $\boldsymbol{\tau}^N$ is the Newtonian stress in the absence of $\boldsymbol{\Pi}$, and $\boldsymbol{\tau}^{NN}$ is the additional Newtonian stress induced by $\boldsymbol{\Pi}$. Consequently, the Newtonian stresslet decomposes as:
\begin{equation}\label{eq:NewtStressletDecomp}
	\hat{\text{\textbf{S}}}(\boldsymbol{\tau}) = \hat{\text{\textbf{S}}}(\boldsymbol{\tau}^N) + \hat{\text{\textbf{S}}}(\boldsymbol{\tau}^{NN}).
\end{equation}
Here, $\hat{\text{\textbf{S}}}(\boldsymbol{\tau}^N)$ corresponds to the Newtonian stresslet (e.g., the Einstein correction $2.5\phi$ for a suspension of spheres \cite{einstein2005neue}), while $\hat{\text{\textbf{S}}}(\boldsymbol{\tau}^{NN})$ represents the Newtonian stresslet induced by non-Newtonian effects. The total contribution from non-Newtonian interactions, defined as:
\begin{equation}
	\hat{\text{\textbf{S}}}_{\text{int}} := \hat{\text{\textbf{S}}}(\boldsymbol{\tau}^{NN}) + \hat{\text{\textbf{S}}}(\boldsymbol{\Pi}),
\end{equation}
is referred to as the {interaction stresslet}. This quantity characterizes how the particle interacts with the microstructure of the fluid, such as polymers in a viscoelastic medium.

Since $\boldsymbol{\tau}^{NN}$ is driven by $\boldsymbol{\Pi}$, the interaction stresslet $\hat{\text{\textbf{S}}}_{\text{int}}$ can be attributed entirely to the non-Newtonian stress. Below, we derive an expression for $\hat{\text{\textbf{S}}}_{\text{int}}$ solely in terms of $\boldsymbol{\Pi}$. Unlike the standard stresslet (defined via a surface integral after applying the divergence theorem to a volume integral \cite{batchelor1970stress}), the interaction stresslet arises from the cumulative effect of the deviation of $\boldsymbol{\Pi}$ from its undisturbed state in the entire fluid volume.

{In the absence of inertia, the mass and momentum equations in \eqref{eq:momentum} are linear in $\mathbf{u}$ and $p$, though nonlinearity may arise in the combined system of these equations along with the constitutive relation for $\begin{bmatrix}\mathbf{u}& p& \boldsymbol{\Pi}\end{bmatrix}$.} Therefore, velocity, pressure and fluid stress can be decomposed into two components:
\begin{equation}
	\mathbf{u} = \mathbf{u}^{N} + \mathbf{u}^{NN}, \quad p = p^{N} + p^{NN}, \quad \boldsymbol{\sigma} = \boldsymbol{\sigma}^{N} + \boldsymbol{\sigma}^{NN},
\end{equation}
with the superscript $N$ representing the purely Newtonian component which satisfy the Newtonian problem governed by:
\begin{equation}
	\nabla\cdot \mathbf{u}^{N}=0,\hspace{0.2in}\nabla\cdot \boldsymbol{\sigma}^{N}=0,\label{eq:NewtonianProblem}
\end{equation}
with $\boldsymbol{\sigma}^N =\boldsymbol{\tau}^N = -p^N\boldsymbol{\delta} + 2\mathbf{e}^N$, and $\mathbf{e}^N = (\nabla \mathbf{u}^N + \nabla \mathbf{u}^{N\text{T}})/2$. Boundary conditions for the Newtonian problem are identical to those for the full problem. Hence the Newtonian problem represents Stokes flow of the originally prescribed conditions around the particle in a Newtonian fluid. The non-Newtonian problem satisfies:
\begin{equation}
	\nabla \cdot \mathbf{u}^{NN} = 0, \quad \nabla \cdot \boldsymbol{\sigma}^{NN} = 0,
\end{equation}
with the non-Newtonian velocity, $\mathbf{u}^{NN}$, and an interaction stress $\boldsymbol{\sigma}^{NN}$ described by:
\begin{equation}
	\boldsymbol{\sigma}^{NN} {= \boldsymbol{\tau}^{NN}+\boldsymbol{\Pi}}= -p^{NN} \boldsymbol{\delta} + 2 \mathbf{e}^{NN} + \boldsymbol{\Pi},
\end{equation}
where $p^{NN}$ is the non-Newtonian pressure, and $\mathbf{e}^{NN}$, the rate of strain field, is defined as $\mathbf{e}^{NN}=\frac{1}{2}(\nabla\mathbf{u}^{NN}+(\nabla\mathbf{u}^{NN})^\text{T})$. The velocity boundary conditions are:
\begin{equation}
	\mathbf{u}^{NN}=0, \text{ at }\textbf{r}=\textbf{r}_p,\text{ and }
\mathbf{u}^{NN}= 0,\text{ as }\textbf{r}\rightarrow \infty.\label{eq:nonNewtProblemBC}
\end{equation}

Rewriting the non-Newtonian problem as:
\begin{equation}
	\nabla \cdot \boldsymbol{\tau}^{NN} = -\nabla \cdot \boldsymbol{\Pi},\label{eq:nonNewtProblem}
\end{equation}
with $\boldsymbol{\tau}^{NN} = \boldsymbol{\tau} - \boldsymbol{\tau}^N$, we observe that the non-Newtonian stress $\boldsymbol{\Pi}$ leads to a body force in the non-Newtonian subproblem. Here, $\boldsymbol{\tau}^{N}$ represents the purely Newtonian stress, absent any non-Newtonian influence and $\boldsymbol{\tau}^{NN}$ the residual Newtonian stress driven by the divergence of $\boldsymbol{\Pi}$.

Beyond calculating $\boldsymbol{\Pi}$ through the non-Newtonian constitutive equation \ref{eq:GeneralConstitutive}, evaluating $\boldsymbol{\tau}^{NN}$ will require the solution (often numerical) of the partial differential equation \eqref{eq:nonNewtProblem}. However, we can express the stresslet generated by $\boldsymbol{\tau}^{NN}$, $\hat{\text{\textbf{S}}}(\boldsymbol{\tau}^{NN})$, directly as a function of $\boldsymbol{\Pi}$, using a generalized reciprocal theorem as considered next. 

Let $\mathbf{v}$ be an auxiliary Stokes velocity field (e.g., extensional deformation at the particle surface that decays to zero at infinity), satisfying:
\begin{align}
	&\boldsymbol{\nabla}\cdot \boldsymbol{\Sigma}=0,\hspace{0.2in}\Sigma_{ijkl}=\delta_{ij}q_{kl}+\frac{\partial {\text{v}}_{jkl}}{\partial r_{i}}+\frac{\partial {\text{v}}_{ikl}}{\partial r_{j}},\hspace{0.2in}
	\boldsymbol{\nabla}\cdot \mathbf{v}=0\label{eq:AuxEqns}\\
	\text{v}_{ijk}&=\text{v}_{surf,jkl}=\frac{1}{2}(\delta_{ik}r_{j}+\delta_{ij}r_{k})-\frac{1}{3}\delta_{kj}r_i,\text{ on }\textbf{r}=\textbf{r}_p,\text{ and }\mathbf{v}=\mathbf{v}_{\infty}\rightarrow 0,\text{ as }\textbf{r}\rightarrow \infty.\label{eq:AuxEqnsBCs}
\end{align}
The tensor $\text{v}_{ijk} \langle e \rangle_{jk}$ represents the Stokes velocity disturbance induced by a particle in an imposed straining flow with rate-of-strain tensor $\langle e \rangle_{jk}$. {Similarly, $\Sigma_{ijkl} \langle e \rangle_{kl}$ and $q_{kl}\langle e \rangle_{kl}$ are the fluid stress and pressure.} For certain geometries, such as spheroids, $\text{v}_{ijk}$ can be obtained analytically \cite{chwang1975hydromechanics,dabade2015effects}. For a spherical particle, the solution is:
\begin{equation}
	\text{v}_{ijk}=\Big(\frac{5}{2r^5}-\frac{5}{2r^7}\Big)r_ir_jr_k+\frac{1}{2r^5}\Big(r_j\delta_{ik}+r_k\delta_{ij}\Big)+\Big(\frac{1}{2r^5}-\frac{5}{6r^3}\Big)r_i\delta_{jk}.\label{eq:AuxillaryVelocity}
\end{equation}
For more complex shapes, $\text{v}_{ijk}$ can be computed numerically by solving the Stokes equations once and reused across a range of non-Newtonian parameter regimes as $\mathbf{v}$ depends only on the particle geometry. 

Consider the following integral identity:
\begin{equation}
	\int_{V_f}dV\text{  } \frac{\partial}{\partial {r}_i}({\tau}^{NN}_{ij} \text{v}_{jkl}-{\Sigma}_{ijkl} {u}^{NN}_j) = \int_{V_f}dV\text{  } \Bigg(\frac{\partial {\tau}^{NN}_{ij}}{\partial {r}_i} \text{v}_{jkl}+{\tau}^{NN}_{ij}\frac{\partial \text{v}_{jkl}}{\partial {r}_{{i}}}-\frac{\partial \Sigma_{ijkl}}{\partial {r}_i}{u}^{NN}_j-\frac{\partial u_j^{NN}}{\partial {r}_i}\Sigma_{ijkl}\Bigg).
\end{equation}
Using equations \eqref{eq:nonNewtProblem} and \eqref{eq:AuxEqns}, this simplifies to:
\begin{equation}
	\int_{V_f}dV\text{  } \frac{\partial}{\partial {r}_i}({\tau}^{NN}_{ij} \text{v}_{jkl}-{\Sigma}_{ijkl} {u}^{NN}_j) = -\int_{V_f}dV\text{  } \frac{\partial {\Pi}_{ij}}{\partial {r}_i} \text{v}_{jkl}.
\end{equation}
{Here, 
	$\tau^{NN}_{ij}\,\dfrac{\partial v_{jkl}}{\partial r_i}
	=\dfrac{\partial u^{NN}_j}{\partial r_i}\,\Sigma_{ijkl}$
	follows from the symmetry of $\tau^{NN}_{ij}$ and $\Sigma_{ijkl}$ in the indices $i$ and $j$, together with the divergence-free nature of the velocities $v_{jkl}$ and $u^{NN}_j$ (i.e., $\partial v_{ikl}/\partial r_i=0$ and $\partial u^{NN}_i/\partial r_i=0$). Specifically,
	\begin{equation}
	\tau^{NN}_{ij}\,\frac{\partial v_{jkl}}{\partial r_i}
	=\Big(\frac{\partial u^{NN}_i}{\partial r_j}+\frac{\partial u^{NN}_j}{\partial r_i}\Big)\frac{\partial v_{jkl}}{\partial r_i},
	\end{equation}
	and, using $\Sigma_{ijkl}=\delta_{ij}\,q_{kl}+\dfrac{\partial v_{jkl}}{\partial r_i}+\dfrac{\partial v_{ikl}}{\partial r_j}$,
	\begin{equation}
	\frac{\partial u^{NN}_j}{\partial r_i}\,\Sigma_{ijkl}
	=\frac{\partial u^{NN}_j}{\partial r_i}\Big(\frac{\partial v_{jkl}}{\partial r_i}+\frac{\partial v_{ikl}}{\partial r_j}\Big)
	=\frac{\partial u^{NN}_j}{\partial r_i}\frac{\partial v_{jkl}}{\partial r_i}
	+\frac{\partial u^{NN}_i}{\partial r_j}\frac{\partial v_{jkl}}{\partial r_i}
	=\Big(\frac{\partial u^{NN}_i}{\partial r_j}+\frac{\partial u^{NN}_j}{\partial r_i}\Big)\frac{\partial v_{jkl}}{\partial r_i}.
	\end{equation}
	Hence the two expressions are equal.}
Applying the divergence theorem yields:
\begin{equation}
	\int_{V_f}dV\text{  } \frac{\partial {\Pi}_{ij}}{\partial x_i} \text{v}_{jkl}=\int_{\mathbf{r}=\mathbf{r}_p}dA\text{  } n_i({\tau}^{NN}_{ij} \text{v}_{jkl}-{\Sigma}_{ijkl} {u}^{NN}_j)-\int_{A_\infty}dA\text{  } n_i({\tau}^{NN}_{ij} \text{v}_{jkl}-{\Sigma}_{ijkl} {u}^{NN}_j).
\end{equation}
Using the boundary conditions in \eqref{eq:nonNewtProblemBC} and \eqref{eq:AuxEqnsBCs}, the second surface integral vanishes, yielding:
\begin{equation}
	\int_{V_f}dV\text{  } \frac{\partial {\Pi}_{ij}}{\partial x_i} \text{v}_{jkl}=\int_{\mathbf{r}=\mathbf{r}_p}dA\text{  } n_i{\tau}^{NN}_{ij} \text{v}_{jkl}.\label{eq:StressletIdentification}
\end{equation}
At the particle surface, where $\mathbf{v} = \mathbf{v}_{\text{surf}}$, the stresslet (equation \eqref{eq:Stresslet1}) due to any stress tensor $\boldsymbol{\sigma}$ can also be written as:
\begin{equation}\label{eq:AlternativeStresslet}
	\hat{\text{\textbf{S}}}(\boldsymbol{\sigma})=\int_{\mathbf{r}=\mathbf{r}_p} dA \text{ }\text{\textbf{n}}\cdot\boldsymbol{\sigma}\cdot  \mathbf{v}_{surf}.
\end{equation}
Therefore, the right-hand side of \eqref{eq:StressletIdentification} is the stresslet due to $\boldsymbol{\tau}^{NN}$:
\begin{equation}
	\hat{\text{\textbf{S}}}(\boldsymbol{\tau}^{NN})=\int_{V_f}dV\text{  } (\nabla \cdot \boldsymbol{\Pi})\cdot \mathbf{v}=\int_{V_f}dV\text{  } (\nabla \cdot (\boldsymbol{\Pi}-  \boldsymbol{\Pi}^{U}))\cdot \mathbf{v}, \label{eq:VolStressGen1}
\end{equation}
where $\boldsymbol{\Pi}^U = \lim_{r \to \infty} \boldsymbol{\Pi}$ is spatially uniform, implying $\nabla \cdot \boldsymbol{\Pi}^U = 0$. Applying the chain rule and the divergence theorem, we obtain:
\begin{align}\label{eq:S1_tau2}\begin{split}
		\hat{\text{\textbf{S}}}(\boldsymbol{\tau}^{NN})=&-\int_{\mathbf{r}=\mathbf{r}_p} dA \text{\textbf{ n}}\cdot(\boldsymbol{\Pi}-\boldsymbol{\Pi}^{U})\cdot \mathbf{v} + \int_{\mathbf{r}\rightarrow\infty}dA \text{\textbf{ n}}\cdot(\boldsymbol{\Pi}-\boldsymbol{\Pi}^{U})\cdot \mathbf{v}\\&-\int_{V_f}dV\text{  } 
		(\boldsymbol{\Pi}-\boldsymbol{\Pi}^{U})^T:\nabla\mathbf{v}.
\end{split}\end{align}
Since $\mathbf{v} \sim r^{-2}$ as $r \to \infty$, the second surface integral vanishes. The first term corresponds to the difference between the stresslets due to $\boldsymbol{\Pi}$ and $\boldsymbol{\Pi}^U$ (via \eqref{eq:AlternativeStresslet}), leading to:
\begin{align}\label{eq:S1_tau3}\begin{split}
		\hat{\text{\textbf{S}}}(\boldsymbol{\tau}^{NN}) = \hat{\text{\textbf{S}}}(\boldsymbol{\Pi}^{U}) - \hat{\text{\textbf{S}}}(\boldsymbol{\Pi}) - \int_{V_f}dV\text{  } (\boldsymbol{\Pi} - \boldsymbol{\Pi}^{U})^T:\nabla\mathbf{v}.
\end{split}\end{align}
Thus, the interaction stresslet becomes:
\begin{align}\begin{split}
		\hat{\text{\textbf{S}}}(\boldsymbol{\sigma}^{NN}) &= \hat{\text{\textbf{S}}}(\boldsymbol{\tau}^{NN}) + \hat{\text{\textbf{S}}}(\boldsymbol{\Pi}) = \hat{\text{\textbf{S}}}(\boldsymbol{\Pi}^{U}) - \int_{V_f}dV\text{  } (\boldsymbol{\Pi} - \boldsymbol{\Pi}^{U})^T:\nabla\mathbf{v}.
	\end{split}\label{eq:totalStressDecomp}\end{align}
This expression enables interpretation of the interaction stresslet in terms of flow geometry. The first term,
\begin{equation}
	\hat{\text{\textbf{S}}}(\boldsymbol{\Pi}^{U}) = V_{particle} \left(\frac{\hat{\boldsymbol{\Pi}}^{{U}} + (\hat{\boldsymbol{\Pi}}^{{U}})^T}{2}\right),
\end{equation}
represents the undisturbed stresslet for a particle of volume $V_{\text{particle}}$. It can also be thought of as a stresslet on an equivalent phantom-particle surface placed in the far-field and experiencing far-field or undisturbed velocity gradients and stress. The second term,
\begin{equation}\label{eq:Svolumediverg1}
V_{particle}	\hat{\text{\textbf{S}}}_\text{volume}(\boldsymbol{\Pi}, \boldsymbol{\Pi}^U) = -\int_{V_f}dV\text{  } (\boldsymbol{\Pi} - \boldsymbol{\Pi}^{U})^T:\nabla\mathbf{v},
\end{equation}
is the volumetric contribution from the deviation of $\boldsymbol{\Pi}$ from its far-field value.

Finally, the ensemble-averaged stress in the suspension (from equation \eqref{eq:SuspensionStress2}) for particles with volume fraction $\phi = n V_{\text{particle}}$ is:
\begin{equation}\label{eq:SuspensionStress4}
	\langle\hat{\boldsymbol{\sigma}}\rangle = 2\langle\mathbf{e}\rangle + \frac{\phi}{V_{particle}}\hat{\text{\textbf{S}}}(\boldsymbol{\sigma}^N) + \hat{\boldsymbol{\Pi}}^U + \phi\Big[\frac{\hat{\boldsymbol{\Pi}}^{{U}} + (\hat{\boldsymbol{\Pi}}^{{U}})^T}{2} + 	\hat{\text{\textbf{S}}}_\text{volume}(\boldsymbol{\Pi}, \boldsymbol{\Pi}^U)+ \text{PINNS}(\boldsymbol{\Pi},\boldsymbol{\Pi}^U,\boldsymbol{\Pi}^L)\Big],
\end{equation}
The terms on the right-hand side are respectively the imposed rate-of-strain tensor, the Newtonian stresslet, the undisturbed non-Newtonian stress, and the total interaction stress. Importantly, all non-Newtonian contributions are expressed in terms of $\boldsymbol{\Pi}$, making this formulation broadly applicable to any fluid whose constitutive relation can be written in the general form of equation \eqref{eq:GeneralConstitutive}. As shown in the following section, under suitable conditions, $\boldsymbol{\Pi}$ can be approximated using the Newtonian velocity field, thereby avoiding the computational cost associated with solving nonlinear partial differential equations.

\subsection{Suspension stress in a weakly non-Newtonian fluid}\label{sec:SuspensionStressWeakNonNewtonian}
The constitutive equations generally require numerical discretization alongside the fluid momentum and mass conservation equations to determine both the Newtonian and non-Newtonian stresses prior to evaluating the rheological quantities introduced above. However, when the non-Newtonian stress is significantly smaller than the Newtonian stress, due to weak non-Newtonian behavior arising from, for instance, low polymer concentration in viscoelastic fluids, slight anisotropy in liquid crystals, or a dilute concentration of spheroids in a Newtonian solvent, a regular perturbation expansion can be performed in the relevant flow variables:
\begin{align}
	\boldsymbol{\sigma} &= \boldsymbol{\sigma}^{(0)} + \epsilon \boldsymbol{\sigma}^{(1)} + \mathcal{O}(\epsilon^2), \\
	\boldsymbol{\tau} &= \boldsymbol{\tau}^{0} + \epsilon \boldsymbol{\tau}^{(1)} + \mathcal{O}(\epsilon^2), \\
	\boldsymbol{\Pi} &= \epsilon \boldsymbol{\Pi}^{(1)} + \mathcal{O}(\epsilon^2), \\
	\mathbf{u} &= \mathbf{u}^{(0)} + \epsilon \mathbf{u}^{(1)} + \mathcal{O}(\epsilon^2), \\
	p &= p^{(0)} + \epsilon p^{(1)} + \mathcal{O}(\epsilon^2),
\end{align}
where $\epsilon$ is a small physical parameter characterizing the strength of the non-Newtonian effects. We assume that $\boldsymbol{\Pi}^{(0)} = 0$, i.e., the non-Newtonian stress vanishes in the limit of a purely Newtonian fluid ($\epsilon = 0$). 

The decomposition of the momentum equation into Newtonian and non-Newtonian components from equations \eqref{eq:NewtonianProblem} to \eqref{eq:nonNewtProblem} in the previous section indicates that $\boldsymbol{\sigma}^{(0)}=\boldsymbol{\tau}^{({0})}=\boldsymbol{\sigma}^{N}=\boldsymbol{\tau}^{N}$ truncates at the leading order and $\boldsymbol{\sigma}^{NN}$ starts at $\mathcal{O}(\epsilon)$, i.e., $\boldsymbol{\sigma}^{NN}=\epsilon \boldsymbol{\sigma}^{(1)} + \mathcal{O}(\epsilon^2)$. Since the leading order term in $\boldsymbol{\Pi}$ appears at $\mathcal{O}(\epsilon)$, the leading-order non-Newtonian stress is influenced solely by the leading-order velocity field, $\mathbf{u}^{(0)}$, corresponding to Newtonian Stokes flow around the particle. The leading-order constitutive equation thus becomes:
\begin{equation}
	\mathbf{H}(\boldsymbol{\Pi}^{(1)}, \mathbf{u}^{(0)}) = 0.
\end{equation}

This Stokes velocity field, $\mathbf{u}^{(0)}$, around an isolated particle, whether available analytically or computed once numerically, can be used to evaluate the first-order non-Newtonian stress. As shown by Koch et al. \cite{koch2016stress} in the context of viscoelastic suspensions, $\boldsymbol{\Pi}^{(1)}$ can be computed efficiently by integrating along the streamlines of the Newtonian velocity field using the method of characteristics. This approach applies to a broad class of non-Newtonian constitutive models in which $\boldsymbol{\Pi}$ (or related quantities) is transported along the flow streamlines by convection. The leading-order constitutive equation reduces to an ordinary differential equation along these streamlines or characterstics. Koch et al. \cite{koch2016stress} introduced this method in the context of dilute polymeric suspensions, and we generalize it here to encompass arbitrary non-Newtonian fluids. Furthermore, Koch et al. \cite{koch2016stress} also demonstrated that a generalized reciprocal theorem could be used to compute the first-order stresslet in the solvent without explicitly solving the momentum equations at $\mathcal{O}(\epsilon)$, provided the polymer stress is known. The derivation in the previous section extends this utility of the reciprocal theorem to any non-Newtonian fluid, even at finite polymer concentration.

In the present framework, once the regular perturbation is performed and $\boldsymbol{\Pi}^{(1)}$ is known, it can be directly substituted into equation \eqref{eq:SuspensionStress4} to yield the ensemble-averaged suspension stress up to $\mathcal{O}(\epsilon)$:
\begin{align}\label{eq:SuspensionStresseorderepsilon3}\begin{split}
		\langle\hat{\boldsymbol{\sigma}}\rangle = & 2\langle\mathbf{e}\rangle + \frac{\phi}{V_{particle}}\hat{\text{\textbf{S}}}(\boldsymbol{\tau}^{(0)}) + \epsilon \hat{\boldsymbol{\Pi}}^{(1U)} + \epsilon \phi\left[ \frac{\hat{\boldsymbol{\Pi}}^{{(1U)}} + (\hat{\boldsymbol{\Pi}}^{{(1U)}})^T}{2} \right. \\
		& \left. + 	\hat{\text{\textbf{S}}}_\text{volume}(\boldsymbol{\Pi}^{(1)}, \boldsymbol{\Pi}^{(1U)})+ \text{PINNS}(\boldsymbol{\Pi}^{(1)},\boldsymbol{\Pi}^{(1U)},\boldsymbol{\Pi}^{(1L)})\right].
\end{split}\end{align}
The terms in square brackets represent the total stress contribution due to the interaction between particles and the non-Newtonian component of the fluid at leading order. Importantly, once the Newtonian velocity field is available, either analytically or from a single numerical computation, all quantities in this equation can be evaluated without any further discretization of partial differential equations.

Thus, the rheology of a weakly non-Newtonian suspension is completely described up to $\mathcal{O}(\epsilon)$ using the Newtonian flow solution alone. We demonstrate the utility of this approach through two illustrative examples in the following section.

\section{Examples: Sphere suspensions in weakly non-Newtonian fluids}\label{sec:examples}
In this section, we explore the rheology of dilute suspensions of spherical particles in two distinct weakly non-Newtonian fluids: a spheroidal fluid and a nematic liquid crystal with slight anisotropy.

{Suspensions of particles in nematic liquid crystals (LCs) exhibit tunable rheological properties due to the anisotropic coupling between suspended inclusions and the LC microstructure~\cite{smalyukh2018liquid}. Even a small concentration of particles can significantly alter the flow response. For example, silica microparticles with homeotropic anchoring dispersed in 8CB liquid crystal transform the material’s behavior from shear thinning to shear thickening, driven by alignment of the nematic director field and reorganization of suspended particles under flow~\cite{sahoo2017rheological}. Such sensitivity of LC rheology to particle addition has implications for LC-based sensors, actuators, and additive manufacturing processes like direct ink writing (DIW), where flow during extrusion is critical to performance. Motivated by this, Section~\ref{sec:LCs} investigates how a small number of rigid spheres modify the shear and extensional rheology of a weakly anisotropic LC, under the simplifying assumption that the director field remains uniform and undisturbed by particles.}

We begin by investigating the rheological effect of particle–microstructure interactions in a spheroidal fluid. A spheroidal fluid is defined as a dilute suspension of small, non-Brownian spheroids in a Newtonian solvent, with the spheroids forming the internal microstructure of the fluid. Such fluids offer a platform for tailoring rheological properties and provide a model system for understanding suspension behavior in more complex non-Newtonian media. Importantly, they also serve as a simplified analog for viscoelastic polymer solutions. For example, Kamdar \textit{et al.}~\cite{kamdar2022colloidal} showed that the enhanced motility of flagellated bacteria in rigid-rod suspensions qualitatively mirrors that observed in polymeric solutions. {In the current context of particle suspensions, spheroidal fluids not only enhance our understanding of the rheology of particle-laden fluids but may also offer an industrially viable alternative to polymer-based suspensions.}

\subsection{Steady state extensional rheology of particle suspensions in a spheroidal fluid}\label{sec:SphdFluid}
We consider an additional suspension of much larger spherical particles within the spheroidal fluid, such that the sphere radius greatly exceeds the major axis of the spheroids. Due to their small size and low concentration, the microstructural spheroids do not significantly disturb the flow streamlines at leading order. Consequently, we assume the spheroid centroids follow the streamlines, and the local flow field around each spheroid is approximately linear, with leading-order corrections in $l_{spheroid}/a_{sphere}$, where $l_{spheroid}$ and $a_{sphere}$ are the major length of the spheroids and the radius of the spheres. The velocity gradients experienced by the spheroids reflect the disturbance induced by the larger suspended spheres. Higher-order corrections to the spheroid velocity, such as those described by Faxén's laws, are of order $\mathcal{O}(l_{spheroid}/a_{sphere})^2$ and are neglected in this analysis. Also, the larger corrections in a thin layer of the size proportional to the length scale of the microstructural spheroids along the suspended spheres is ignored.

The orientation vector of a spheroid, $\mathbf{d}$, evolves with time $s$ according to \cite{kim2013microhydrodynamics}:
\begin{equation}\label{eq:OrientationEvolution}
	\frac{d}{ds}{\mathbf{d}}=\boldsymbol{\Omega}\cdot \mathbf{d}+B(\mathbf{e}\cdot\mathbf{d}-(\mathbf{e}:\mathbf{dd})\mathbf{d}), \quad B=\frac{\kappa^2-1}{\kappa^2+1},
\end{equation}
where $\boldsymbol{\Omega} = (\nabla\mathbf{u} - \nabla \mathbf{u}^T)/2$ and $\mathbf{e} = (\nabla\mathbf{u} + \nabla \mathbf{u}^T)/2$ are the local vorticity and rate-of-strain tensors, respectively, experienced by a spheroid of aspect ratio $\kappa$.

We consider a steady uniaxial extensional flow where the imposed rate-of-strain and vorticity tensors are given by:
\begin{equation}\label{eq:UndisturbedStrainRate}
	\langle\mathbf{e}\rangle_{ij} = \delta_{i1}\delta_{j1} - \frac{1}{2}(\delta_{i2}\delta_{j2} + \delta_{i3}\delta_{j3}), \quad \langle\boldsymbol{\Omega}\rangle = 0.
\end{equation}
The orientation vector $\mathbf{d}$ in Cartesian and spherical coordinates is:
\begin{equation}
	\mathbf{d} = \begin{bmatrix} d_1 \ d_2 \ d_3 \end{bmatrix} = \begin{bmatrix} \cos(\theta) \ \sin(\theta)\cos(\phi) \ \sin(\theta)\sin(\phi) \end{bmatrix},
\end{equation}
where $d_1$ is the component along the extensional axis, and $\theta$ and $\phi$ are the polar and azimuthal angles.
The undisturbed $d_1$ component is obtained as follows:
\begin{equation}\label{eq:Undisturbedorientationsolution}
	\frac{d}{ds} d_1^U = 1.5 B d_1^U (1 - (d_1^U)^2) \quad \Rightarrow \quad d_1^U(s) = \frac{\exp(1.5 B s)}{\sqrt{\exp(3B s) - 1 + (d_1^U(0))^{-2}}},
\end{equation}
and the azimuthal angle $\phi^U$ remains constant in the absence of disturbances from the suspended spheres:
\begin{equation}
	\frac{d\phi^U}{ds} = 0.
\end{equation}

In steady state, prolate spheroids ($\kappa > 1$, $B > 0$) align with the extensional axis, yielding $\mathbf{d}^U_{\kappa>1} = [1, 0, 0]$. Oblate spheroids ($\kappa < 1$, $B < 0$) align their face normal in the compressional plane, resulting in:
\begin{equation}\label{eq:StartingOrientations}
	\mathbf{d}^U_{\kappa<1} = [0, d_2^U, d_3^U], \quad d_2^U = \cos{\phi^U}, \quad d_3^U = \sin{\phi^U}, \quad \phi^U \in [0, \pi].
\end{equation}
where $\phi^U$ is the initial (and subsequent) azimuthal angle.

Once the flow is disturbed by the spheres, the local orientation of the spheroids varies spatially. In the steady-state, we solve equation \eqref{eq:OrientationEvolution} along streamlines of the flow field, using $s$ as the streamwise coordinate. The tensors $\mathbf{e}(\mathbf{r})$ and $\boldsymbol{\Omega}(\mathbf{r})$ are computed from the analytical Stokes flow solution around a sphere in an unbounded extensional flow. To leading order in spheroid concentration $\epsilon$, the velocity and pressure fields are:
\begin{eqnarray}
	&u_i^{(0)}=\begin{cases}
		\langle{\Omega}\rangle_{ij}r_j+\langle\text{e}\rangle_{ij}r_j+\frac{5}{2}\Big(\frac{1}{r^7}-\frac{1}{r^5}\Big)\langle\text{e}\rangle_{jk}r_jr_kr_i-\frac{1}{r^5}\langle\text{e}\rangle_{ji}r_j,& r\ge 1,\\
		\langle{\Omega}\rangle_{ij}r_j,& r<1,
	\end{cases}\label{eq:u_field}\\
	&p^{(0)}=-\frac{5}{r^5}\langle\text{e}\rangle_{jk}r_jr_k , \hspace{2.2in}r\ge 1,
\end{eqnarray}
This velocity field allows us to evaluate the local velocity gradient tensor at each point and compute the orientation dynamics numerically using MATLAB’s ode15s solver. The initial condition for solving the ordinary differential equation \eqref{eq:OrientationEvolution} for the spheroidal orientation field is the undisturbed orientation from equation \eqref{eq:StartingOrientations}.

The non-Newtonian stress generated by the spheroidal microstructure is given by the following constitutive equation:
\begin{equation}\label{eq:SpheroidalFluidRheology}
	\epsilon \boldsymbol{\Pi}^{(1)} = \epsilon \Big[ 4A_H (\mathbf{e} : \langle \mathbf{dddd} \rangle) + 4B_H (\mathbf{e} \cdot \langle \mathbf{dd} \rangle + \langle \mathbf{dd} \rangle \cdot \mathbf{e} - \tfrac{2}{3} \boldsymbol{\delta} \mathbf{e} : \langle \mathbf{dd} \rangle) + 2C_H \mathbf{e} \Big],
\end{equation}
where $A_H$, $B_H$, and $C_H$ are geometry-dependent functions of aspect ratio $\kappa$ as tabulated in \cite{kim2013microhydrodynamics}, and $\langle \cdot \rangle$ denotes an orientation average. This equation models the first-order (in $\epsilon$) stress contribution from a dilute suspension of aligned spheroids in a Newtonian fluid under deformation rate $\mathbf{e}$.

{As considered in equations \eqref{eq:Undisturbedorientationsolution} to \eqref{eq:StartingOrientations}, while there is a single steady-state undisturbed orientation of a prolate spheroid, an oblate spheroid admits many possible undisturbed orientations given by $d_1^U = 0$, $\phi^U \in [0, \pi]$. This degeneracy does not affect the undisturbed stress, which depends only on the rate of strain tensor averaged over a uniform distribution in $\phi^U$:}
\begin{align}\label{eq:SpheroidalFluidRheologyUndisturbed}
	\hat{\boldsymbol{\Pi}}^{(1U)}_{\kappa>1}= \Big\{ \frac{8}{3}A_H +\frac{16}{3} B_H+2 C_H   \Big\} \langle\mathbf{e}\rangle, \hspace{0.2in}	\hat{\boldsymbol{\Pi}}^{(1U)}_{\kappa<1}= \Big\{ 2 A_H +\frac{4}{3} B_H+2 C_H   \Big\} \langle\mathbf{e}\rangle.
\end{align}

To quantify the effect of flow disturbances caused by the spheres on the microstructural orientation and resulting stress, we require the leading order (in $\mathcal{O}(\epsilon)$) undisturbed  ($\hat{\boldsymbol{\Pi}}^{(1U)}$), linearized ($\hat{\boldsymbol{\Pi}}^{(1L)}$) and total ($\hat{\boldsymbol{\Pi}}^{(1)}$) non-Newtonian stress. This requires linearizing the orientation dynamics about the undisturbed state. Let $\mathbf{d} = \mathbf{d}^U + \mathbf{d}^{(L)}$, where $|\mathbf{d}^{(L)}| \ll | \mathbf{d}^U|$. Substituting into equation \eqref{eq:OrientationEvolution} and retaining terms linear in $\mathbf{d}^{(L)}$ and the disturbance in strain rate $\mathbf{e}' = \mathbf{e} - \langle\mathbf{e}\rangle$, we obtain the linearized evolution equation:
\begin{equation}\label{eq:OrientationEvolutionLinear}
	\frac{d}{ds} \mathbf{d}^{(L)} = \boldsymbol{\Omega} \cdot \mathbf{d}^U + B \left[ \mathbf{e}' \cdot \mathbf{d}^U + \langle\mathbf{e}\rangle \cdot \mathbf{d}^{(L)} - \langle\mathbf{e}\rangle : ( \mathbf{d}^{(L)}\mathbf{d}^U\mathbf{d}^U + \mathbf{d}^U\mathbf{d}^{(L)}\mathbf{d}^U + \mathbf{d}^U\mathbf{d}^U\mathbf{d}^{(L)} ) - \mathbf{e}' : \mathbf{d}^U\mathbf{d}^U\mathbf{d}^U \right].
\end{equation}

This linear response $\mathbf{d}^{(L)}$ is used to compute the linearized correction to the stress tensor, $\boldsymbol{\Pi}^{(1L)}$. Substituting $\mathbf{d} = \mathbf{d}^U + \mathbf{d}^{(L)}$ into equation \eqref{eq:SpheroidalFluidRheology}, we obtain:
\begin{align}\label{eq:SpheroidalFluidRheologyLinear}
	\begin{split}
		&\boldsymbol{\Pi}^{(1L)}=\\& 4A_H \langle\mathbf{e}\rangle:\langle(\mathbf{d}^{U}\mathbf{d}^{U}\mathbf{d}^{U}\mathbf{d}^{(L)} +\mathbf{d}^{U}\mathbf{d}^{U}\mathbf{d}^{(L)}\mathbf{d}^{U}+\mathbf{d}^{U}\mathbf{d}^{(L)}\mathbf{d}^{U}\mathbf{d}^{U}\\&+ \mathbf{d}^{(L)}\mathbf{d}^{U}\mathbf{d}^{U}\mathbf{d}^{U})\rangle_\text{orient.}+4A_H \mathbf{e}':\mathbf{d}^{U}\mathbf{d}^{U}\mathbf{d}^{U}\mathbf{d}^{U}\rangle_\text{orient.}\\&+4 B_H [\langle\mathbf{e}\rangle\cdot\langle(  \mathbf{d}^{(L)}\mathbf{d}^{U}+ \mathbf{d}^{U}\mathbf{d}^{(L)})\rangle_\text{orient.}+\mathbf{e}'\cdot  \langle\mathbf{d}^{U}\mathbf{d}^{U}\rangle_\text{orient.}\\&+\langle(  \mathbf{d}^{(L)}\mathbf{d}^{U}+ \mathbf{d}^{U}\mathbf{d}^{(L)})\rangle_\text{orient.}\cdot\langle\mathbf{e}\rangle+ \langle\mathbf{d}^{U}\mathbf{d}^{U}\rangle_\text{orient.}\cdot\mathbf{e}'
		-\frac{2}{3}\boldsymbol{\delta}\hspace{0.05in}(\mathbf{e}':\langle\mathbf{d}^{U}\mathbf{d}^{U}\rangle_\text{orient.}\\&+\langle\mathbf{e}\rangle:\langle(\mathbf{d}^{(L)}\mathbf{d}^{U}+\mathbf{d}^{U}\mathbf{d}^{(L)})\rangle_\text{orient.})] +2 C_H \mathbf{e}'.
\end{split}\end{align}
By substituting equations \eqref{eq:Undisturbedorientationsolution}, \eqref{eq:StartingOrientations}, \eqref{eq:SpheroidalFluidRheology}, \eqref{eq:SpheroidalFluidRheologyUndisturbed}, and \eqref{eq:SpheroidalFluidRheologyLinear}, along with the numerically evaluated orientation distributions $\mathbf{d}$ and $\mathbf{d}^L$, into \eqref{eq:PINNS} and \eqref{eq:Svolumediverg1}, we obtain the required terms to evaluate the suspension stress given by equation \eqref{eq:SuspensionStresseorderepsilon3}. Similar to the undisturbed stress shown in equation \eqref{eq:SpheroidalFluidRheologyUndisturbed}, for oblate spheroids, an orientation average for the interaction stress is taken over a uniform distribution of the phase angle, $\phi^U \in [0, \pi]$ with $d_1^U = 0$. The non-Newtonian contributions, arising from the undisturbed and interaction stresses, are discussed below.

\begin{figure}
	\centering
	\subfloat{\includegraphics[width=0.47\textwidth]{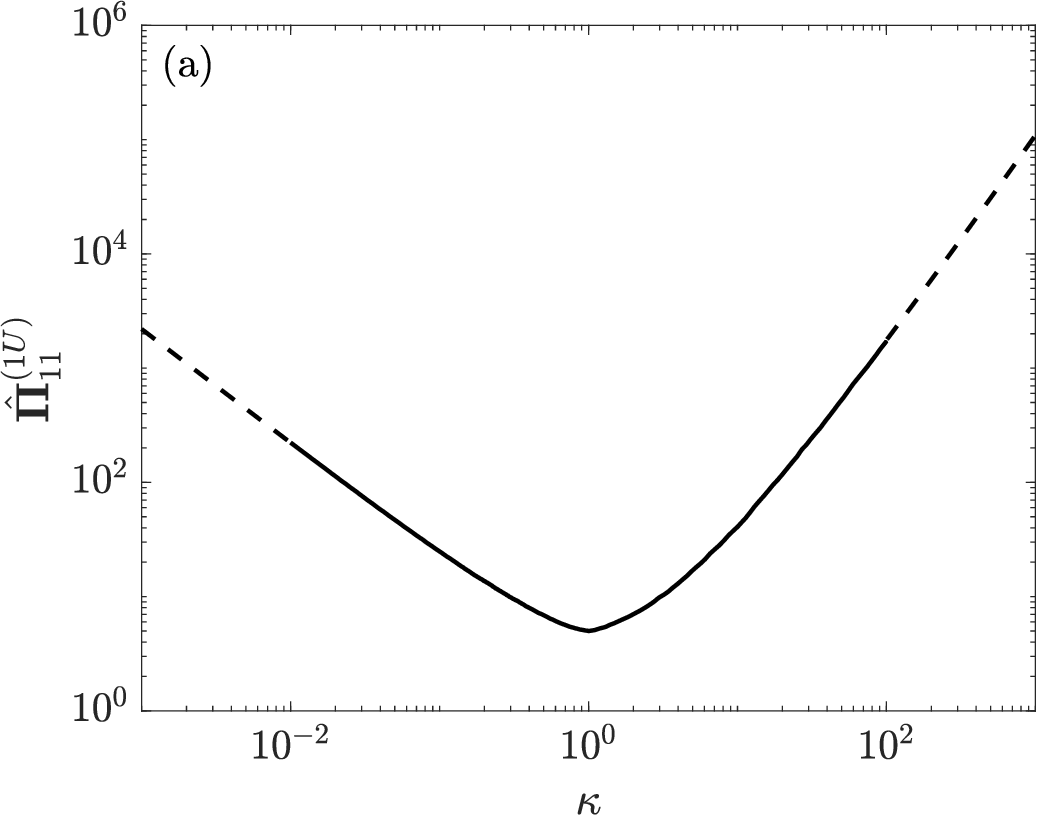}\label{fig:UndisturbedStress.eps}}\hfill
	\subfloat{\includegraphics[width=0.51\textwidth]{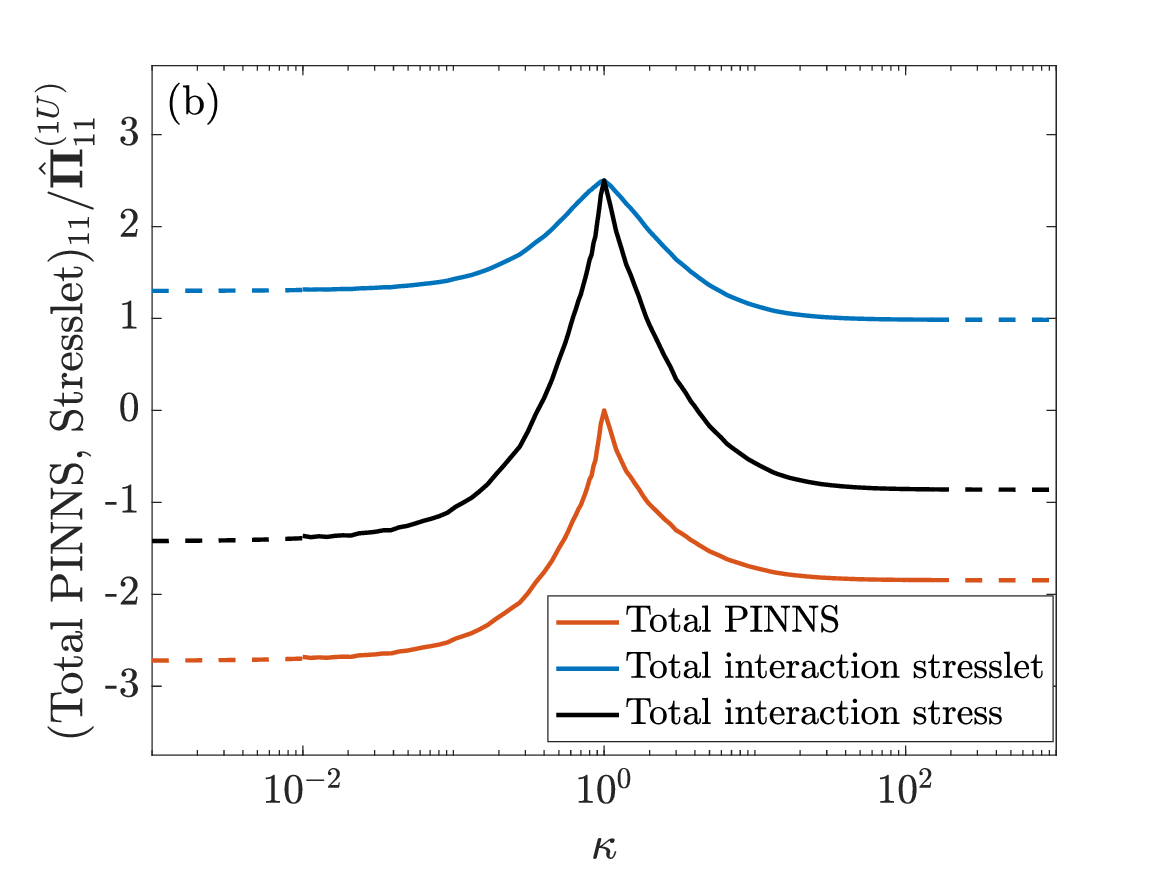}\label{fig:ProlateOblateSpheroidalFluidTotal}}
	\caption {(a) Undisturbed stress (computed using stress functions from \cite{kim2013microhydrodynamics}) and (b) particle-spheroid interaction stress in a suspension of spheres in a dilute spheroidal fluid across different aspect ratios $\kappa$. {Dashed lines indicate the limiting behavior in the prolate ($\kappa \rightarrow \infty$) and oblate ($\kappa \rightarrow 0$) regimes.} \label{fig:ProlateOblateSpheroidalFluid}}
\end{figure}

The undisturbed non-Newtonian stress, $\hat{\boldsymbol{\Pi}}^{(1U)}$, of a spheroidal fluid as a function of the microstructural aspect ratio $\kappa$ is shown in figure \ref{fig:UndisturbedStress.eps}. For both prolate and oblate spheroidal fluids, the magnitudes of the undisturbed and interaction stresses increase as $\kappa$ deviates from unity.
Figure \ref{fig:ProlateOblateSpheroidalFluidTotal} displays the total interaction stress normalized by $\phi\hat{\boldsymbol{\Pi}}^{(1U)}$, along with its decomposition into the total PINNS and the interaction stresslet. For $\kappa = 1$, this normalized value is 2.5, corresponding to a suspension of spheres in a Newtonian fluid, for which the PINNS contribution vanishes. As $\kappa$ deviates from unity, both the normalized PINNS and interaction stresslet decrease. In the asymptotic limits, the normalized stresslet approaches 1.0 for slender fibers ($\kappa \gg 1$) and 1.3 for thin discs ($\kappa \ll 1$), while the PINNS approaches about –1.84 and –2.7, respectively. Thus, the total suspension stress becomes negative for sufficiently large or small $\kappa$, ($\kappa\lessapprox 0.36 $ and $\kappa\gtrapprox 3.9$) indicating that adding spheres to the spheroidal fluid reduces the extensional stress. This stress reduction mirrors that observed in dilute suspensions of spheres in polymeric (viscoelastic) fluids \citep{sharma2023steady}.

To understand the origin of this behavior, we decompose the suspension stress into contributions from the $A_H$, $B_H$, and $C_H$ terms in equation \eqref{eq:SpheroidalFluidRheology}. Specifically, we define $\boldsymbol{\Pi}^{(1)}_{AH}=4\mathbf{e}:\langle \mathbf{dddd} \rangle_\text{orient.}  $, $\boldsymbol{\Pi}^{(1)}_{BH} =4 [\mathbf{e}\cdot \langle \mathbf{dd}\rangle_\text{orient.}+\langle \mathbf{dd}\rangle_\text{orient.}\cdot\mathbf{e}-\frac{2}{3}\boldsymbol{\delta}\mathbf{e}:\langle\mathbf{dd}\rangle_\text{orient.}]$ and $\boldsymbol{\Pi}^{(1)}_{CH} =2\mathbf{e} $ which allow us to rewrite the total non-Newtonian stress (equation \eqref{eq:SpheroidalFluidRheology} is) as,
\begin{equation}\label{eq:SpheroidalFluidRheology 3}
	\boldsymbol{\Pi}^{(1)}=A_H \boldsymbol{\Pi}^{(1)}_{AH} +B_H \boldsymbol{\Pi}^{(1)}_{BH}  +C_H \boldsymbol{\Pi}^{(1)}_{CH}.
\end{equation}
The $C_H$-related stresses ($\boldsymbol{\Pi}^{(1)}_{CH}$, $\boldsymbol{\Pi}^{(1U)}_{CH}$ and $\boldsymbol{\Pi}^{(1L)}_{CH}$) are independent of $\kappa$ and yield: PINNS$(\boldsymbol{\Pi}^{(1)}_{CH},\boldsymbol{\Pi}^{(1U)}_{CH},\boldsymbol{\Pi}^{(1L)}_{CH})=0$ and $\hat{\text{\textbf{S}}}_\text{volume}^{(1)}(\boldsymbol{\Pi}^{(1)}_{CH},\boldsymbol{\Pi}^{(1U)}_{CH})=3 C_H$. Hence, the $\kappa$-dependence and the observed reduction in stress of spheroidal fluid upon adding spheres originates from $\boldsymbol{\Pi}^{(1)}_{AH}$ and $\boldsymbol{\Pi}^{(1)}_{BH}$. We now examine the PINNS and interaction stresslet components arising from these two terms, i.e., PINNS$(\boldsymbol{\Pi}^{(1)}_{AH})$, PINNS$(\boldsymbol{\Pi}^{(1)}_{BH})$, $\hat{\text{\textbf{S}}}_\text{volume}^{(1)}(\boldsymbol{\Pi}^{(1)}_{AH},\boldsymbol{\Pi}^{(1U)}_{AH})$ and $\hat{\text{\textbf{S}}}_\text{volume}^{(1)}(\boldsymbol{\Pi}^{(1)}_{BH},\boldsymbol{\Pi}^{(1U)}_{BH})$. 

\begin{figure}
	\centering
	\subfloat{\includegraphics[width=0.49\textwidth]{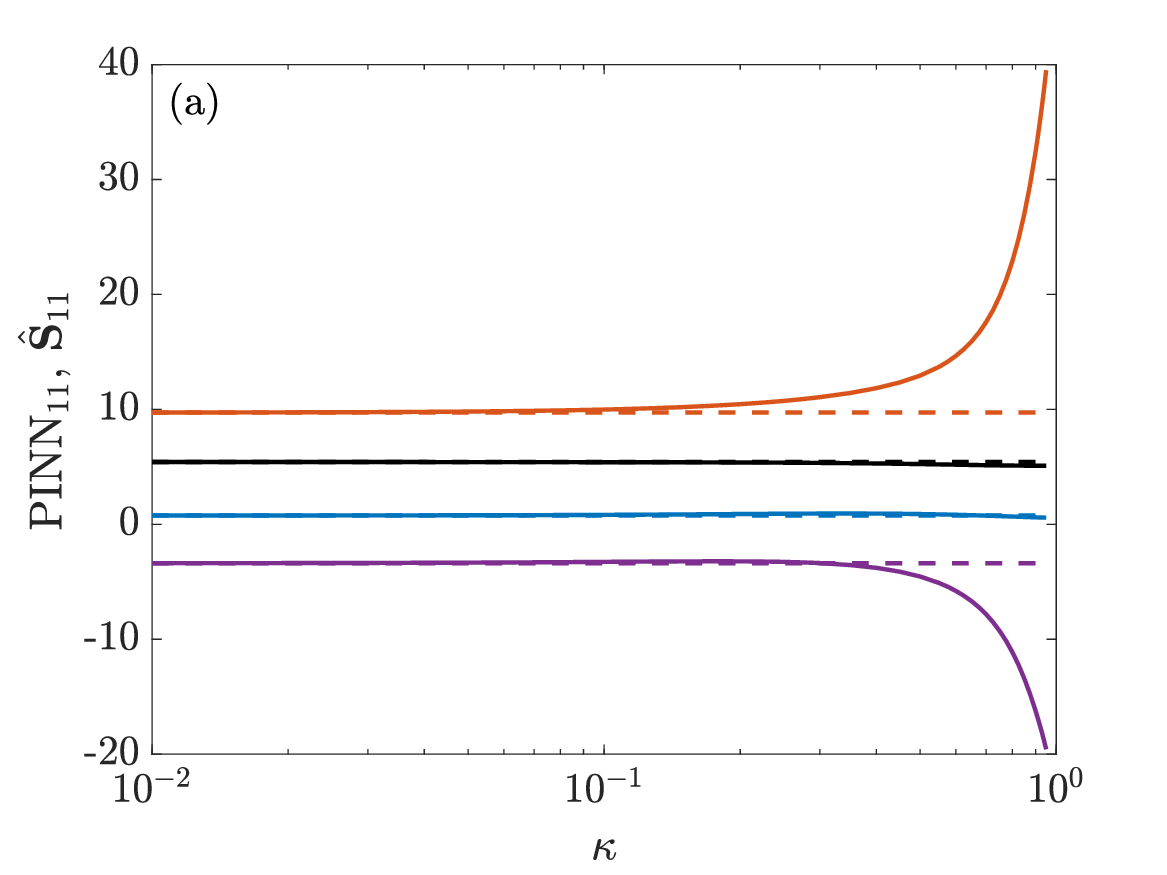}\label{fig:OblateSpheroidalFluidsplit}}\hfill
	\subfloat{\includegraphics[width=0.49\textwidth]{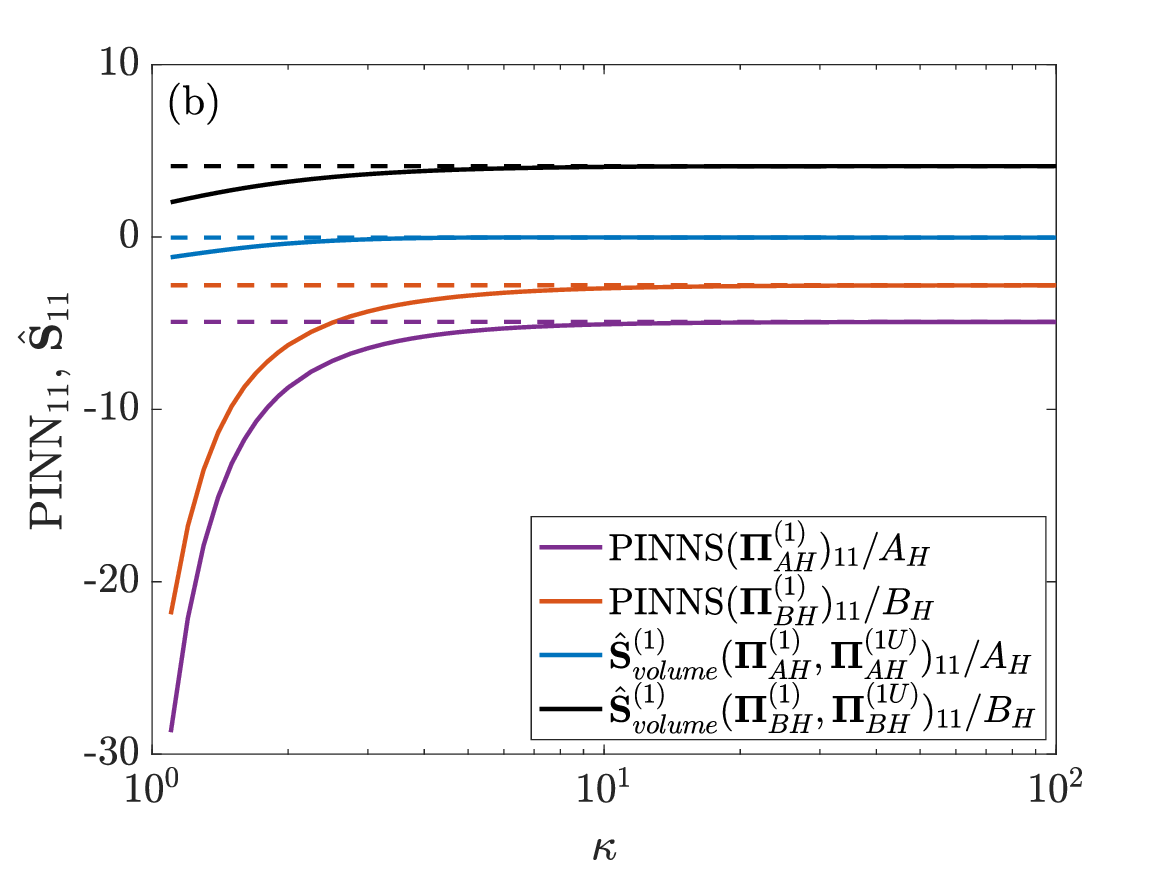}\label{fig:SpheroidalFluidsplit}}
	\caption {Normalized components of particle-spheroidal interaction stress in a suspension of spheres in dilute (a) oblate and (b) prolate spheroidal fluids as a function of $\kappa$. Dashed lines represent asymptotic limits. Both panels share the same legend.\label{fig:ProlateOblateSpheroidalFluidSplit}}
\end{figure}

Figure \ref{fig:ProlateOblateSpheroidalFluidSplit} shows the PINNS and volumetric interaction stresses, normalized by their corresponding undisturbed components, as functions of $\kappa$ for prolate and oblate fluids. The divergence in $\text{PINNS}(\boldsymbol{\Pi}^{(1)}_{AH})/A_H$ and $\text{PINNS}(\boldsymbol{\Pi}^{(1)}_{BH})/B_H$ near $\kappa=1$ arises from the faster decay of $A_H$ and $B_H$ compared to the numerator terms, which tend to zero more slowly.
For prolate fluids, normalized PINNS is always negative. In oblate fluids, normalized PINNS due to $\boldsymbol{\Pi}^{(1)}_{AH}$ is negative while that due to $\boldsymbol{\Pi}^{(1)}_{BH}$ is positive. The signs of $A_H$, $B_H$, and $C_H$ differ between the two types: for oblate fluids, $A_H, C_H > 0$, $B_H < 0$; for prolate fluids, all three are positive \cite{kim2013microhydrodynamics}. Consequently, both $\boldsymbol{\Pi}^{(1)}_{AH}$ and $\boldsymbol{\Pi}^{(1)}_{BH}$ yield negative PINNS for all $\kappa \ne 1$.

In the asymptotic limits ($\kappa \gg 1$ or $\kappa \ll 1$), $A_H \gg B_H, C_H$ \cite{kim2013microhydrodynamics}, and the stress is dominated by the $A_H$ contribution. At $\kappa = 1$, $A_H = B_H = 0$ and $C_H = 5$, yielding $\hat{\boldsymbol{\Pi}}^{(1U)} \rightarrow 10 \langle \mathbf{e} \rangle$. For $\kappa < 1$, the volumetric stresslet from $A_H$ is positive, and that from $B_H$ is negative. Alternatively, for $\kappa > 1$, $\hat{\text{\textbf{S}}}_\text{volume}^{(1)}(\boldsymbol{\Pi}^{(1)}_{AH},\boldsymbol{\Pi}^{(1U)}_{AH})$ is initially negative but vanishes for $\kappa \gtrsim 2.5$, while $\hat{\text{\textbf{S}}}_\text{volume}^{(1)}(\boldsymbol{\Pi}^{(1)}_{BH},\boldsymbol{\Pi}^{(1U)}_{BH})$ remains positive. In the slender fiber limit ($\kappa \rightarrow \infty$), the dominant contributions are:
\begin{align}\begin{split}&\lim\limits_{\kappa\rightarrow\infty}\text{Total Interaction Stresslet}\approx\hat{\boldsymbol{\Pi}}^{(1U)}\approx\frac{8}{3}A_H\langle\mathbf{e}\rangle,\\& \lim\limits_{\kappa\rightarrow\infty}\text{Total PINNS}\approx\text{PINNS}(\boldsymbol{\Pi}^{(1)}_{AH})\approx-1.84 \hat{\boldsymbol{\Pi}}^{(1U)}\approx-4.9A_H\langle\mathbf{e}\rangle,\end{split}\label{eq:LargeARProlate}\end{align}
In the disc-like limit ($\kappa \rightarrow 0$), the values are:
\begin{equation}\lim\limits_{\kappa\rightarrow 0}\text{Total Interaction Stresslet}=1.3 \hat{\boldsymbol{\Pi}}^{(1U)}, \lim\limits_{\kappa\rightarrow 0}\text{Total PINNS}=-2.7 \hat{\boldsymbol{\Pi}}^{(1U)}.\end{equation} 

\begin{figure}
	\centering	
	\subfloat{\includegraphics[width=0.33\textwidth]{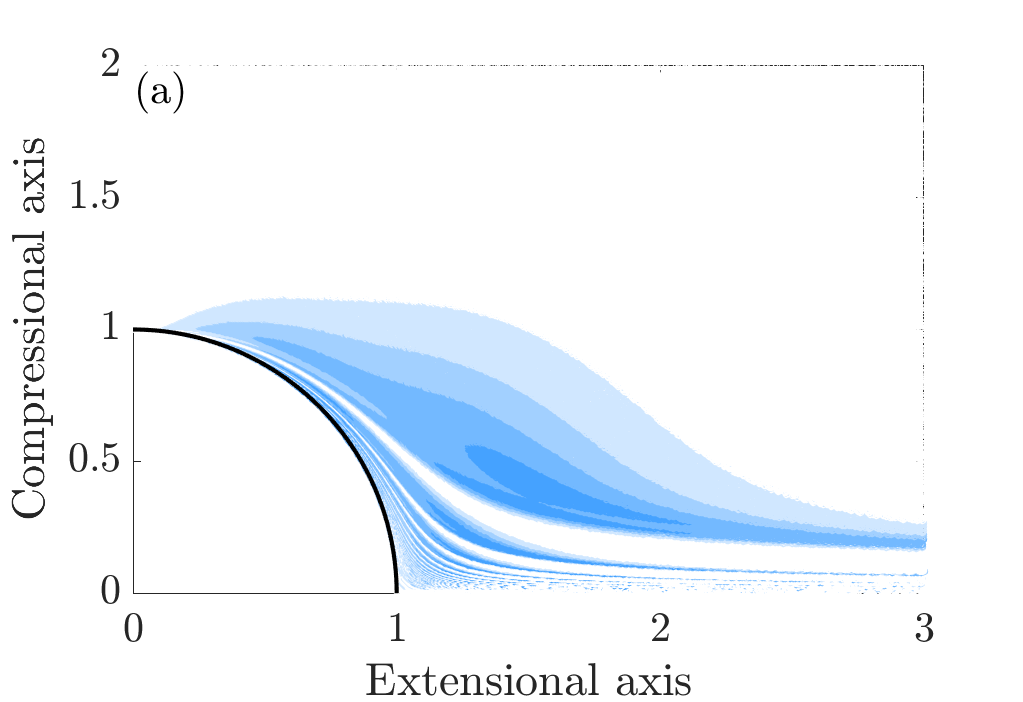}\label{fig:kappa2stress.eps}}\hfill
	\subfloat{\includegraphics[width=0.33\textwidth]{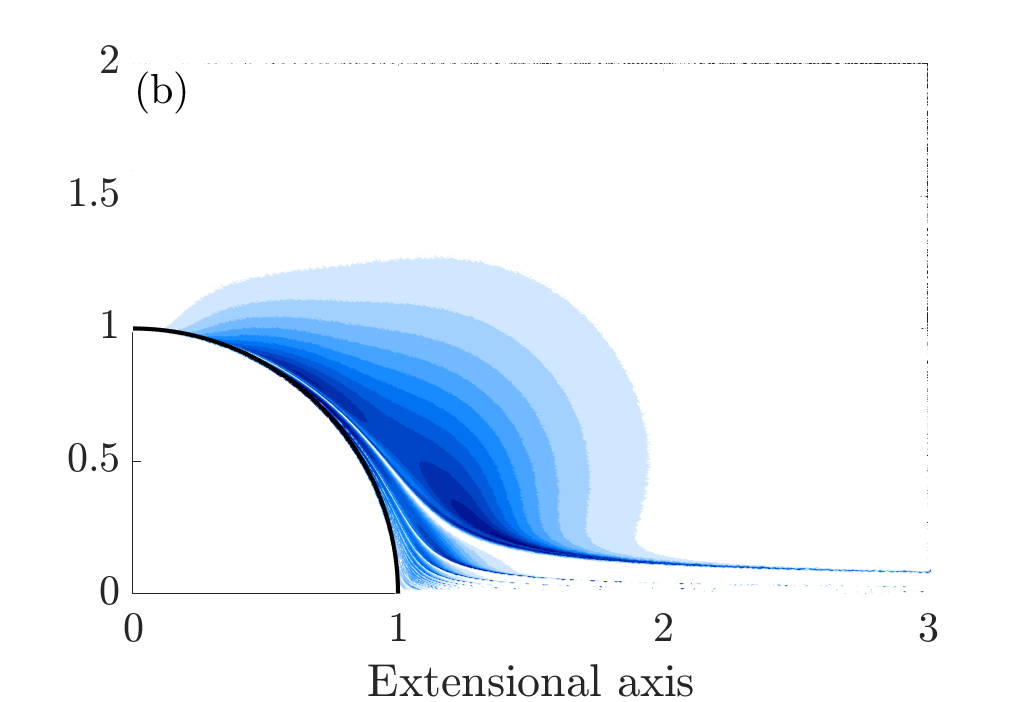}\label{fig:kappa5stress.eps}}\hfill	\subfloat{\includegraphics[width=0.33\textwidth]{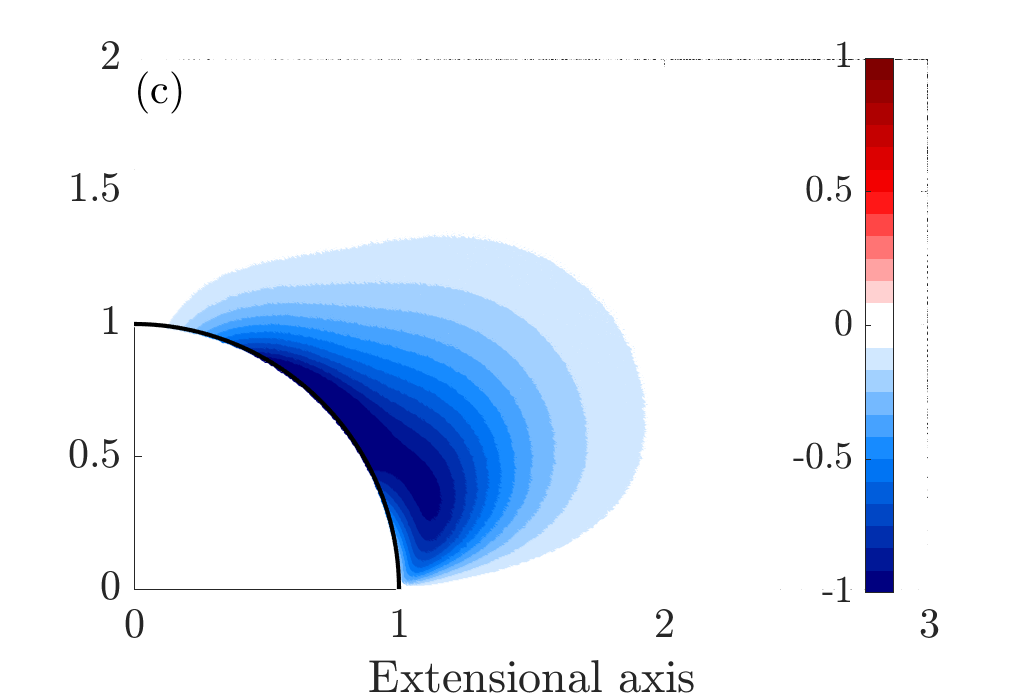}\label{fig:kappainfStess.eps}}
		\caption{Normalized extensional component of the nonlinear stress, $\hat{\Pi}^{(1NL)}_{11} / \hat{\Pi}^{(1U)}_{11}$, for prolate spheroidal fluids with (a) $\kappa = 2$, (b) $\kappa = 5$, and (c) $\kappa \rightarrow \infty$.\label{fig:NLStressProlate}}
\end{figure}

\begin{figure}
	\centering	
	\subfloat{\includegraphics[width=0.33\textwidth]{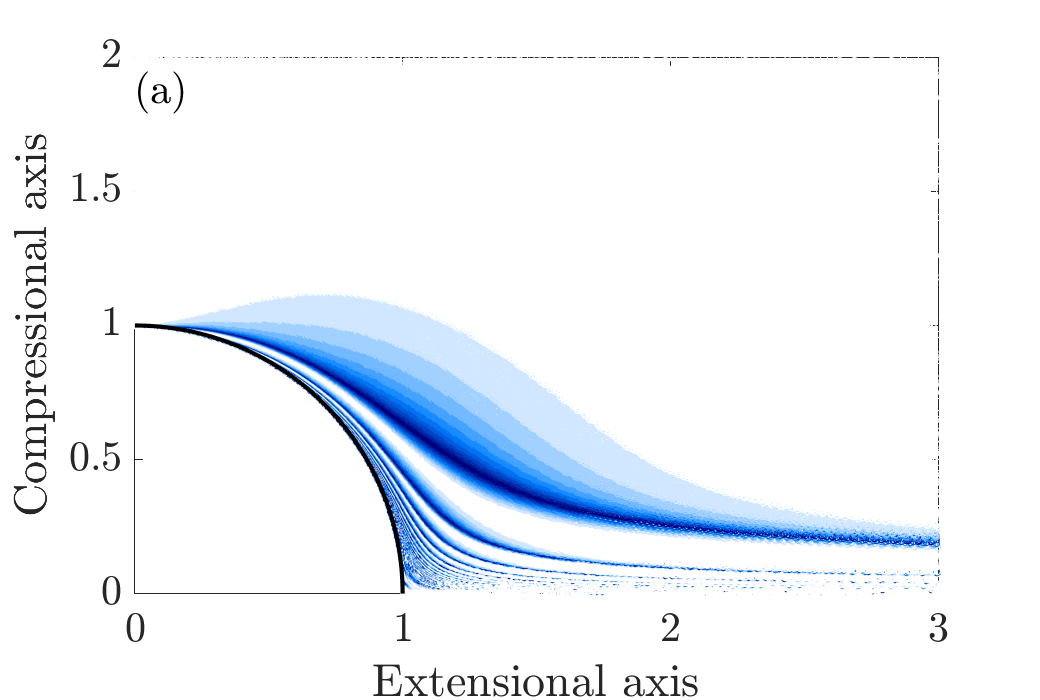}\label{fig:kappa0.eps}}
	\subfloat{\includegraphics[width=0.33\textwidth]{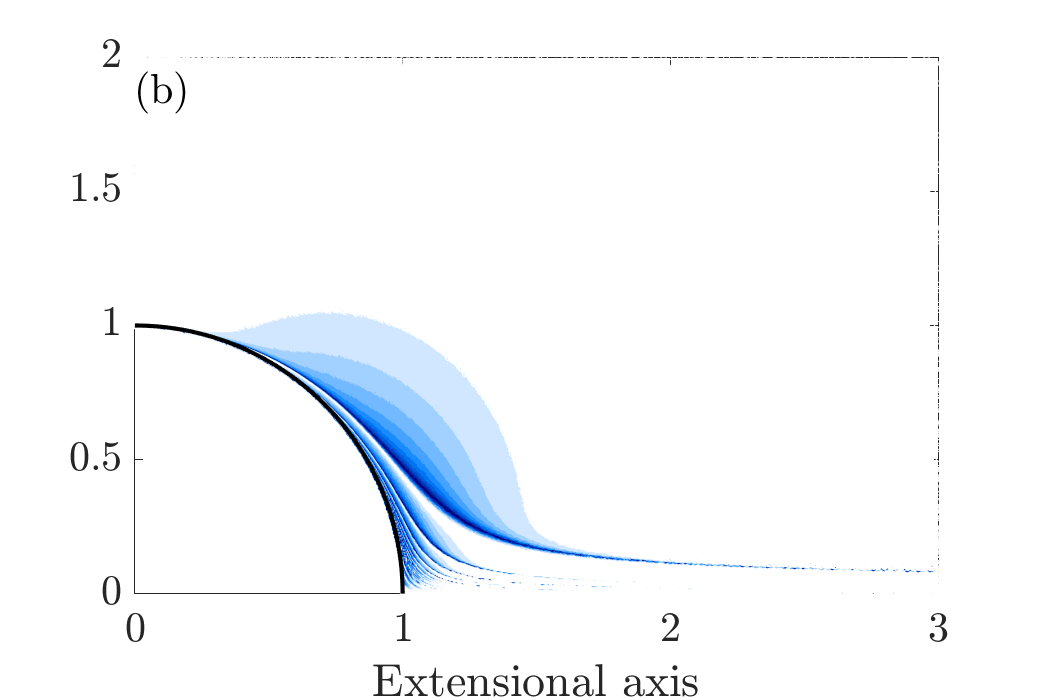}\label{fig:kappap1.eps}}
	\subfloat{\includegraphics[width=0.33\textwidth]{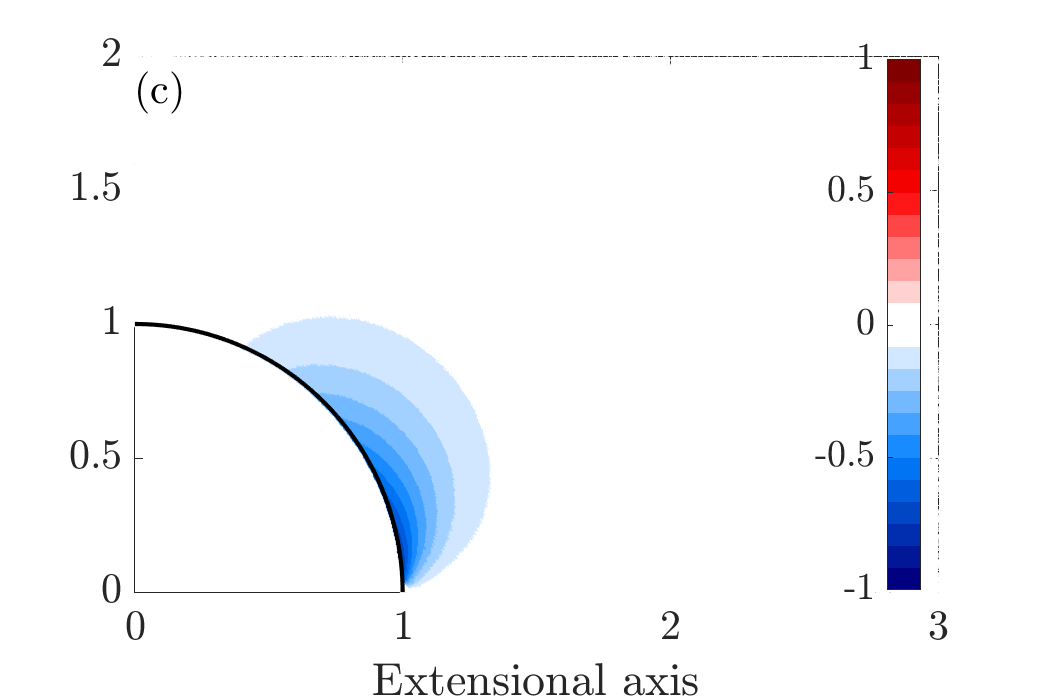}\label{fig:kappap3.eps}}
	\caption{Misalignment of the spheroid orientation along the extensional axis, quantified by $\mathbf{d} \cdot [1, 0, 0]^T - 1$, for prolate spheroidal fluids with (a) $\kappa = 2$, (b) $\kappa = 5$, and (c) $\kappa \rightarrow \infty$.\label{fig:OrientationsProlateSpheroids}}
\end{figure}

\begin{figure}
	\centering	
	\subfloat{\includegraphics[width=0.49\textwidth]{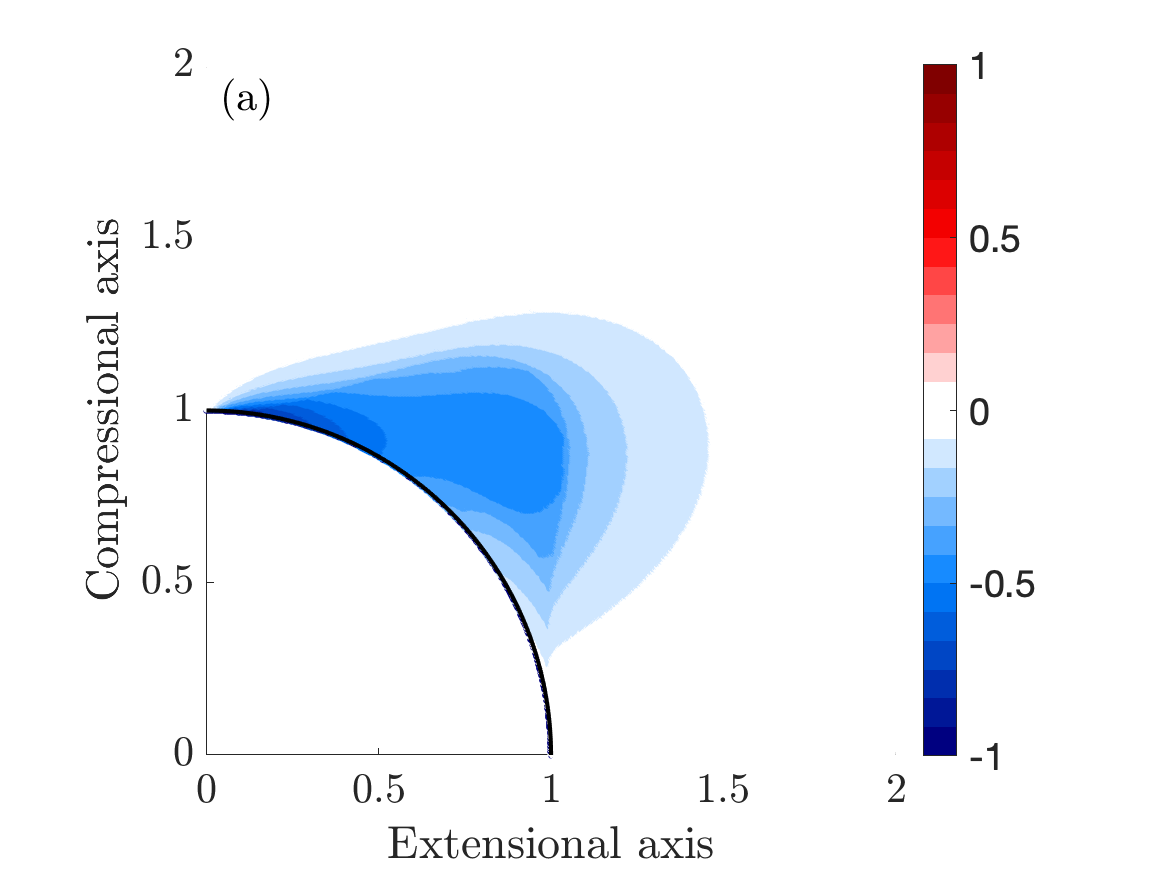}\label{fig:VelocityGradientEigenvReallalignment}}
	\subfloat{\includegraphics[width=0.4\textwidth]{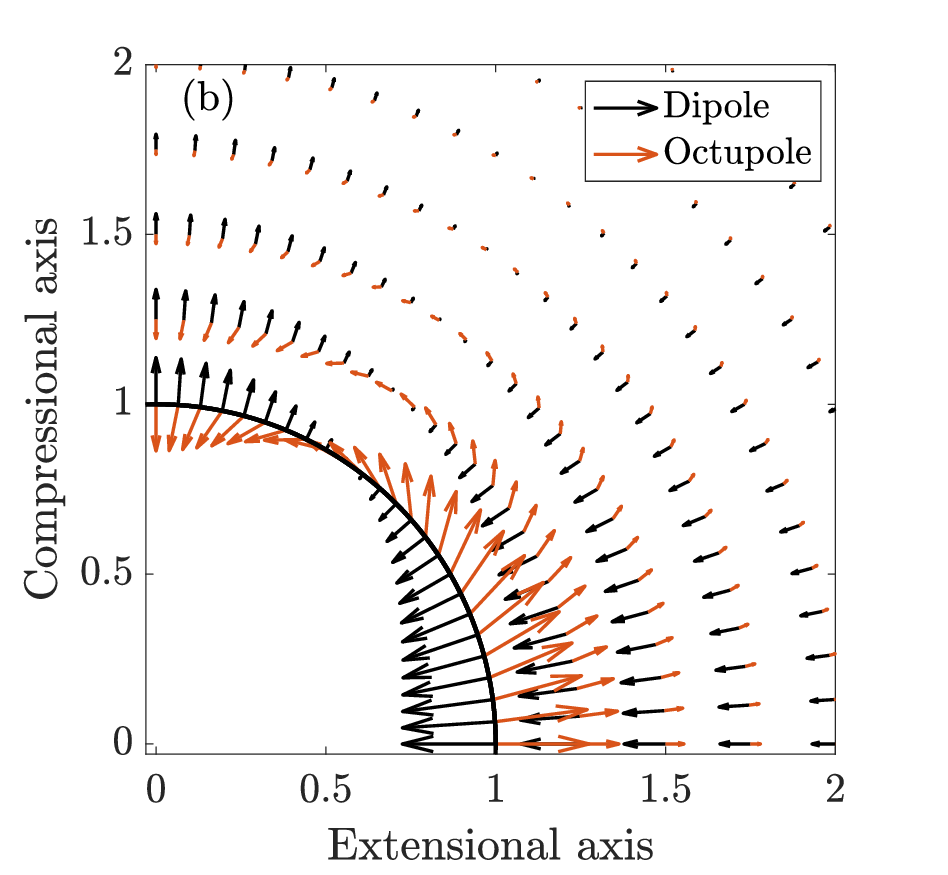}\label{fig:Multipole}}
		\caption{(a) {Misalignment of the local and undisturbed flows' extensional axis, quantified as one minus the projection of the local extensional direction onto that of the imposed flow}. (b) Dipole and octupole disturbance velocity fields induced by the sphere. The dominant contribution to orientation misalignment arises from the dipole component.\label{fig:VelKinematics}}
\end{figure}

To elucidate the mechanism behind the negative PINNS and the resulting net negative interaction stress, we present in figure \ref{fig:NLStressProlate} the spatial distribution of the extensional component of the nonlinear non-Newtonian stress, normalized by the corresponding undisturbed value, i.e., $\hat{\Pi}^{(1NL)}_{11} / \hat{\Pi}^{(1U)}_{11}$, for prolate spheroidal fluids with $\kappa = 2$, $\kappa = 5$, and $\kappa \rightarrow \infty$. In the far field, the spheroids align with the extensional axis of the imposed flow, which coincides with the $x$-axis in figure \ref{fig:NLStressProlate}. To visualize deviations from this alignment in the disturbed flow, we plot $\mathbf{d} \cdot [1, 0, 0]^T - 1$ in figure \ref{fig:OrientationsProlateSpheroids}. The non-white regions in this figure, indicating misalignment between the local and undisturbed spheroidal orientation, correlate with regions of finite nonlinear stress disturbance in figure \ref{fig:NLStressProlate}. Thus, the nonlinear stress $\hat{\Pi}^{(1NL)}_{11}$ arises from local orientation deviations relative to the undisturbed state. For large $\kappa$, these deviations are confined to regions near the particle surface, explaining the observed decrease in the volumetric interaction stress $\hat{\mathbf{S}}^{(1)}_\text{volume}$ with increasing aspect ratio.
To understand the physical origin of these orientation deviations, we analyze the local flow kinematics in the leading-order velocity field around a sphere in a Newtonian fluid. Specifically, we examine the alignment of the local extensional axis, defined as the eigenvector corresponding to the largest (real) eigenvalue of the rate-of-strain tensor, with the undisturbed extensional axis. {Figure \ref{fig:VelocityGradientEigenvReallalignment} shows one minus the projection of the local extensional axis onto the extensional axis of undisturbed flow.} The white regions correspond to near-perfect alignment; deviations occur only in a thin region surrounding the sphere. These regions of misalignment coincide with those of finite nonlinear stress and orientation disturbance in figures \ref{fig:NLStressProlate} and \ref{fig:OrientationsProlateSpheroids}. As a prolate spheroid travels along the compressional plane and enters this zone, its major axis reorients before eventually realigning with the imposed extension direction upon exiting. At lower $\kappa$, this misalignment extends further from the sphere because the reorientation time, inversely proportional to $B$ (and hence $\kappa$), is longer, as evident from equations \eqref{eq:OrientationEvolution} and \eqref{eq:Undisturbedorientationsolution}.

{The disturbance velocity field induced by the sphere includes both dipole ($\mathcal{O}(r^{-2})$) and octupole ($\mathcal{O}(r^{-4})$) components, as shown in figure \ref{fig:Multipole} and described by equation \eqref{eq:u_field}. Consider a flow consisting only of the dipole component. Near the extensional axis, the velocity points inward and grows along the direction of motion, stretching fluid elements along the $x$ axis. Near the compressional axis, the velocity points outward and decays, compressing fluid elements in the $y$ direction and extending them in $x$ due to incompressibility. Thus, the dipole-induced local extensional axis remains aligned with the corresponding axis of the imposed flow in both regions, producing no misalignment. In contrast, the octupole velocity field, which generally points opposite to the dipole near the extensional and compressional axes, induces extension perpendicular to that caused by the imposed flow, resulting in complete misalignment. However, the negligible misalignment in these regions seen in figure \ref{fig:VelocityGradientEigenvReallalignment} shows that the dipole effect dominates overall. Significant misalignment occurs only in a region centered around $\tan^{-1}(\sqrt{2}) \approx 54.7^\circ$ from the $x$ axis, where both dipole and octupole disturbances contribute. This is reflected in the negative (blue) region of figure \ref{fig:VelocityGradientEigenvReallalignment}, and leads to deviation in the local orientation of the microstructural spheroids and hence the negative PINNS discussed above.}

A similar mechanism governs the nonlinear stress response for oblate spheroids, where the orientation of the face normal replaces that of the major axis. In these cases (not shown), the regions of orientation misalignment and finite nonlinear stress are analogous to those seen for prolate spheroids, and their spatial extent increases with decreasing flatness (i.e., larger $\kappa$ for oblate shapes), for the same reason that the extent increases at lower $\kappa$ in prolate cases.

{The stress reduction mechanism observed here bears a strong analogy to the localized polymer stress response in the extensional flow of a dilute suspension of spheres in a viscoelastic fluid, as analyzed in our previous work~\citep{sharma2023steady}. In that case, for large polymer relaxation times, the polymer molecules are nearly fully stretched in the undisturbed extensional flow, leading to a large elastic stress that plays a role analogous to the non-Newtonian stress generated by spheroidal microstructure (equation~\eqref{eq:SpheroidalFluidRheology}). As the polymers approach the sphere, the reduced local velocity gradients cause them to relax toward their equilibrium, unstretched state. Upon exiting this zone, they re-extend, resulting in a localized region where the polymer stretch deviates from that in the undisturbed flow. As the Deborah number (the product of the polymer relaxation time and the imposed extension rate) is decreased, while remaining large, the polymer takes longer to recover its fully stretched state, enlarging the region of stretch disturbance. Thus, the Deborah number in viscoelastic fluids plays a role analogous to the spheroid aspect ratio $\kappa$ in a spheroidal fluid.} Quantitatively, in the large-$\kappa$ limit, the total interaction stress at $\mathcal{O}(\phi\epsilon)$ is $-0.84 \hat{\Pi}^{(1U)}$, as seen from equation \eqref{eq:LargeARProlate}. In our study on viscoelastic fluids \citep{sharma2023steady}, we find that the corresponding interaction stress in the large Deborah number limit is $-0.85 \hat{\Pi}^{(1U)}$, where $\hat{\Pi}^{(1U)}$ is the undisturbed stress in the polymeric fluid. This agreement underscores the mechanistic and quantitative parallels between the two systems.

Additionally, the interaction between spherical particles and a spheroidal fluid provides insight into the rheology of bidisperse suspensions in the limit of large particle size contrast. Theoretical work on simple shear flow of bidisperse spheres has shown that the stresslet arising from hydrodynamic interactions between differently sized particles is reduced in the dilute limit~\cite{wagner1994viscosity}. Although this theory applies strictly to dilute suspensions, it qualitatively explains experimental observations in more concentrated systems, where adding smaller spheres to a suspension of larger ones (at a total volume fraction around 0.2) leads to a decrease in viscosity~\cite{goto1982flow}. These results rely on the numerically evaluated resistivity functions of \cite{jeffrey1992calculation} for two unequal spheres in close proximity, combined with the suspension stress formulation of \cite{batchelor1972determination}. In the limit of large radius ratio, this reduction results from the fact that small particles typically do not see large particles within a distance of order of the small particle radius.  Our work treats the smaller particles as a continuum microstructure and captures their collective interaction with larger particles. We demonstrate that a further reduction in \textit{extensional viscosity} can be achieved by introducing particles of a different shape (spheroids), via a distinct mechanism involving a negative particle-induced non-Newtonian stress (PINNS). This shape-mediated effect offers an additional design parameter for tuning suspension rheology and motivates future experiments on the extensional response of bidisperse suspensions with varying particle geometries.

\subsection{Steady state shear and extensional rheology of particle suspensions in liquid crystals}\label{sec:LCs}
Liquid crystals (LCs) represent a state of matter that lies between an isotropic liquid and a crystalline solid~\cite{de1993physics}. Nematic LCs, in particular, are composed of rod-like molecules that exhibit long-range orientational order without long-range positional order~\cite{chandrasekhar_1992}. The resulting stress anisotropy, i.e., direction-dependent mechanical properties, has enabled applications in direct ink writing (DIW) additive manufacturing~\cite{wang2022anisotropic}, soft robotics~\cite{kotikian2019untethered}, energy-absorbing materials~\cite{traugutt2017liquid}, and the study of bacterial motility in anisotropic environments such as extracellular matrices and biofilms~\cite{mushenheim2014using}. A central feature of these applications is the ability to tune the stress–strain response by manipulating molecular orientation or composition. In this section, we demonstrate that the addition of rigid spheres can modify the rheology of a dilute suspension in weakly anisotropic nematic LCs. 

Nematic LCs are characterized by a director field, along which the viscosity typically differs from that in the perpendicular direction. Recent experiments by Chandrasekar \textit{et al.}~\cite{chandrasekar2023micro} employed micrometer-sized ferrofluid droplets to probe the directional viscosity of the LC 8CB (4-cyano-4'-octylbiphenyl) via magnetic manipulation. They observed that the viscosities parallel and perpendicular to the director deviate continuously, without discontinuities, as the temperature is lowered below the isotropic transition point, $T_\mathrm{iso} \approx 38.3\,^\circ\mathrm{C}$. This trend supports the use of weakly anisotropic models near the isotropic–nematic transition. Similar behavior was observed in micellar solutions of cetyltrimethylammonium bromide (CTAB) in water, where increasing micelle concentration led to a smooth increase in anisotropic viscosity. Comparable results were reported for aqueous solutions of the food dye Sunset Yellow. These observations suggest that our theoretical treatment, focused on the limit of weak mechanical anisotropy, provides a quantitatively accurate approximation near the isotropic–nematic transition, where directional stress differences are small but finite.

The stress in a nematic LC can be written as a combination of Newtonian and non-Newtonian components. The non-Newtonian contribution, $\boldsymbol{\Pi}$, is given by the Leslie–Ericksen constitutive relation~\cite{gomez2013flow}:
\begin{equation}\label{eq:constitutiveNLC}
{\Pi}_{ij}=\alpha_1 n_kn_pe_{kp}n_in_j+\alpha_2n_iN_j+\alpha_3n_jN_i+(\alpha_2+\alpha_3)n_jn_ke_{ki}+\alpha_4(n_in_ke_{kj}+n_jn_ke_{ki}),
\end{equation}
where $\alpha_i$ ($i \in [1,4]$) are the Leslie viscosity coefficients, $\mathbf{n}$ is the unit director field representing the local orientation of the nematic phase, {and $\mathbf{N}$ represents the rotation rate of the director relative to the local vorticity of the flow. This rotation rate, which enters the torque balance governing director dynamics, is defined as:}
\begin{equation}
{\mathbf{N}=\frac{\partial \mathbf{n}}{\partial t}+\mathbf{u}\cdot\nabla\mathbf{n}-\frac{1}{2}(\nabla \times \mathbf{u})\times\mathbf{n}.}
\end{equation}
{Under the simplifying assumption of a fixed director field (i.e., neglecting director dynamics), this reduces to}
\begin{equation}
{\mathbf{N} = -\frac{1}{2}(\nabla \times \mathbf{u}) \times \mathbf{n}.}
\end{equation}
{In experimental or practical settings, a uniform director field may be imposed using external electric or magnetic fields, in the absence of suspended particles. However, spherical particles typically induce topological defects in the director field, including dipolar or Saturn-ring~\cite{stark2001physics,ruhwandl1996friction,stark2001stokes,koenig2009single}, due to surface anchoring, which renders $\mathbf{n}$ spatially non-uniform. In this work, for simplicity, we neglect these particle-induced director distortions in order to isolate the influence of fluid velocity disturbances and stresslets caused by the suspended particles on the suspension rheology. Incorporating these distortions in future studies would require solving for a spatially varying director field $\mathbf{n}(\mathbf{x})$ by balancing the elastic and viscous torques. The non-Newtonian stress, $\boldsymbol{\Pi}$, generated by this non-uniform $\mathbf{n}(\mathbf{x})$ can then be integrated into a numerical framework analogous to the spheroidal fluid example presented in the previous section.}

Here, we take $\mathbf{n} = [1, 0, 0]$, such that the non-Newtonian stress reduces to
\begin{equation}
\Pi_{ij}=\alpha_1\delta_{1i}\delta_{1j}\frac{\partial u_1}{\partial x_1}-\alpha_2\delta_{1i}\frac{\partial u_j}{\partial x_1}+\alpha_3\frac{\partial u_1}{\partial x_i}\delta_{1j}+(\alpha_2+\alpha_4)\Big(\delta_{1i}\frac{\partial u_j}{\partial x_1}+\delta_{1j}\frac{\partial u_i}{\partial x_1}+\frac{\partial u_1}{\partial x_i}\delta_{1j}+\frac{\partial u_1}{\partial x_j}\delta_{1i}\Big),\label{eq:TotalLCStress}
\end{equation}
where the term multiplying $\alpha_2+\alpha_4$ is symmetric in $i, j$. Therefore, in the limit $\alpha_1 = \alpha_2 = \alpha_3 = 0$, the non-Newtonian stress is symmetric, as can also be seen from equation~\eqref{eq:constitutiveNLC}. The stresses due to $\alpha_1$ are associated with elongational strain along the director, and those from the symmetric term are less intuitive. The terms $\alpha_3$ and $-\alpha_2$ represent stress–strain anisotropy and bending-resistance anisotropy, respectively, as seen from the vorticity-based analysis in~\cite{gomez2013flow}. In the net non-Newtonian stress (including the symmetric term), we treat each of $\alpha_i, i\in[1,4]$, individually as a small parameter $\epsilon$ in the formulation of section~\ref{sec:SuspensionStressWeakNonNewtonian}, and present results for all four cases together. Throughout, we assume $\alpha_i \ll 1, i\in[1,4]$, and derive the suspension rheology up to first order in these parameters.

{Due to the fixed-director assumption, $\boldsymbol{\Pi}$ depends only linearly on the flow field. This is evident from equation~\eqref{eq:TotalLCStress} (or equation~\eqref{eq:constitutiveNLC}), i.e., although the non-Newtonian stress varies spatially through the velocity gradients, the mapping $\mathbf{u}\mapsto\boldsymbol{\Pi}$ is linear. Therefore
	$\boldsymbol{\Pi}(\langle \mathbf{u} \rangle+\mathbf{u}')=\boldsymbol{\Pi}(\langle \mathbf{u} \rangle)+\boldsymbol{\Pi}(\mathbf{u}')$ exactly, leaving no nonlinear remainder, i.e., $\boldsymbol{\Pi}^{(NL)}=0$. If the director is allowed to respond to the local flow, $\mathbf{n}=\mathbf{n}(\mathbf{u})$, or in the presence of anchoring to the particle surface, $\boldsymbol{\Pi}^{(NL)}$ need not vanish. Physically, because the director is unresponsive to the particle-induced disturbance, there is no nonlinear interaction between the nematic stress and the flow; only the projection of stresses along $\mathbf{n}$ changes. This implies that the particle-induced non-Newtonian stress (PINNS) is zero, and only the Newtonian and interaction stresslets contribute to the ensemble-averaged suspension stress. If the particle effects  $\mathbf{n}$ through director anchoring and elastic nematic stresses at low Ericksen number, a finite PINNS may be expected involving the non-linear coupling between a spatially varying $\mathbf{n'}$ and $\mathbf{u'}$ (or $\mathbf{e'}$).  Furthermore, in the present analysis of weakly anisotropic nematic LCs, to first order in the anisotropy, $\boldsymbol{\Pi}$ is an explicit function of the Newtonian velocity field (set $\mathbf{u}=\mathbf{u}^{(0)}$ in equation~\eqref{eq:TotalLCStress}), and the volumetric stresslet $\hat{\mathbf{S}}_\text{volume}(\boldsymbol{\Pi},\boldsymbol{\Pi}^U)$ can be obtained analytically from equation~\eqref{eq:Svolumediverg1}. We consider extensional flow and shear rheology below.
}

\subsubsection{{Extensional Rheology}}
For uniaxial extensional flow with the extension axis aligned with the director field, the ensemble-averaged stress is
\begin{equation}\label{eq:ExtensionalLC}
	{\langle\hat{\boldsymbol{\sigma}}\rangle_\text{extension}=\Big((2+5\phi)+\frac{2}{3}(\alpha_3-\alpha_2)+\alpha_1\!\left(\frac{2}{3}+\frac{20}{21}\phi\right)+(\alpha_2+\alpha_4)\!\left(\frac{4}{3}+\frac{5}{2}\phi\right)\Big)\langle\mathbf{e}\rangle.}
\end{equation}
Here, the first term in parentheses corresponds to the Newtonian contribution, while the remaining terms represent corrections due to the nematic phase, with and without particles.
{Particle–LC interactions do not contribute through the $\alpha_3$ or $-\alpha_2$ terms in equation~\eqref{eq:TotalLCStress}: for these terms, the volumetric stresslet $\hat{\mathbf{S}}_\text{volume}(\boldsymbol{\Pi},\boldsymbol{\Pi}^U)$ cancels the undisturbed part (not shown). By contrast, the contributions associated with elongational strain along the director ($\alpha_1$) and the symmetric part $(\alpha_2+\alpha_4)$ are enhanced by the presence of the particle. As shown by equation~\eqref{eq:totalStressDecomp}, in addition to the volume/undisturbed split, we also decompose the interaction stresslet into (i) the LC non-Newtonian extra stress $\boldsymbol{\Pi}$ and (ii) the Newtonian disturbance $\boldsymbol{\tau}^{NN}$ (the perturbation to the Newtonian stress induced by $\boldsymbol{\Pi}$). The former is obtained by substituting $\boldsymbol{\sigma}=\boldsymbol{\Pi}$ from equation~\eqref{eq:TotalLCStress} into the surface-integral expression in equation~\eqref{eq:Stresslet1} (or equation~\eqref{eq:AlternativeStresslet}). While the latter would usually require solving the first-order momentum equation, this is unnecessary here because the net interaction stresslet is already known from the volume/undisturbed decomposition. The resulting contributions are}
\begin{eqnarray}
	{\hat{\text{\textbf{S}}}(\boldsymbol{\Pi})_\text{extension}
		=\Big(\frac{6}{7}\alpha_1-\frac{9}{7}\alpha_2+2\alpha_3+\frac{25}{14}(\alpha_2+\alpha_4)\Big)\phi\langle\mathbf{e}\rangle,}\\
	{\hat{\text{\textbf{S}}}(\boldsymbol{\tau}^{NN})_\text{extension}
		=\Big(\frac{2}{21}\alpha_1+\frac{9}{7}\alpha_2-2\alpha_3+\frac{5}{7}(\alpha_2+\alpha_4)\Big)\phi\langle\mathbf{e}\rangle.}
\end{eqnarray}
{As expected, the stresslet contributions from $\boldsymbol{\Pi}$ and $\boldsymbol{\tau}^{NN}$ cancel for the $-\alpha_2$ and $\alpha_3$ anisotropies. For $\alpha_1$ and $(\alpha_2+\alpha_4)$, $\hat{\text{\textbf{S}}}(\boldsymbol{\Pi})_\text{extension}$ dominates over $\hat{\text{\textbf{S}}}(\boldsymbol{\tau}^{NN})_\text{extension}$; thus, the majority of the additional suspension stress due to particles arises from extra nematic traction on the particle surface rather than from the perturbations to the surface velocity gradient and pressure induced by $\boldsymbol{\Pi}$. The particle-free nematic LC has a finite stress due to bending-resistance anisotropy and stress-strain anisotropy ($-\tfrac{2}{3}\alpha_2\langle\mathbf{e}\rangle$ and  $\tfrac{2}{3}\alpha_3\langle\mathbf{e}\rangle$, respectively), and the individual components of the stresslet under either decomposition are nonzero for these anisotropy types. However, these stresslets exactly cancel, leading to no net coupling of these anisotropies with the particle. By contrast, $\alpha_1$ (featuring $\mathbf{n}\mathbf{n}:\mathbf{e}$) and the symmetric $(\alpha_2+\alpha_4)$ projection directly sample the principal stretch along $\mathbf{n}$; adding a particle therefore strengthens fore–aft normal tractions and increases the first moment (stresslet), leading to the positive $\phi$-coefficients for $\alpha_1$ and $(\alpha_2+\alpha_4)$.}
In DIW applications, where the fluid undergoes an extensional flow as it emerges from the nozzle, such modulation of extensional rheology can be used to optimize performance by adjusting particle concentration.

\subsubsection{{Shear Rheology}}
{Since flow near the nozzle walls in DIW is locally simple shear, we examine the effect of particle–nematic interactions on the shear rheology for two representative director orientations. The first of these is time independent in a particle-free simple shear flow of a flow-aligning liquid crystal, while the second is stationary in a particle-free shear flow of any nematic liquid crystal{, i.e., the director aligned along the imposed vorticity direction}. In simple shear flow, with the director aligned along the flow direction, the suspension stress is:}
\begin{equation}\label{eq:Shear1LC}
	\langle \hat{\boldsymbol{\sigma}} \rangle_\text{shear, n$_\text{flow}$} = \left( (2 + 5\phi) + \frac{1}{2} \alpha_3 (2 {+} \phi) + \frac{1}{2} \alpha_2 \phi + \frac{25}{126} \alpha_1 \phi + (\alpha_2 + \alpha_4)\left(1 + \frac{{25}}{12} \phi \right) \right) \langle \mathbf{e} \rangle.
\end{equation}
{For simple shear $\mathbf{u}=(y,0,0)$, the nonzero components of the rate-of-strain tensor are $e_{12}=e_{21}=1/2$, and the vorticity is $\boldsymbol{\omega}=(0,0,1)$. With $\mathbf{n}$ along the flow, the particle-free LC contributions arise through $\alpha_3$ and $(\alpha_2+\alpha_4)$, which appear as the $\phi$-independent terms in equation~\eqref{eq:Shear1LC}. The corresponding interaction stresslets due to $\boldsymbol{\Pi}$ and $\boldsymbol{\tau}^{NN}$ are}
\begin{eqnarray}
	{\hat{\text{\textbf{S}}}(\boldsymbol{\Pi})_{\text{shear, n}_\text{flow}}
		=\left(\alpha_3 - \tfrac{23}{14}\alpha_2 + \tfrac{1}{14}\alpha_1 + \tfrac{39}{28}(\alpha_2+\alpha_4)\right)\phi\langle \mathbf{e} \rangle,}\\
	{\hat{\text{\textbf{S}}}(\boldsymbol{\tau}^{NN})_{\text{shear, n}_\text{flow}}
		=\left(-\tfrac{1}{2}\alpha_3 + \tfrac{15}{7}\alpha_2 + \tfrac{8}{63}\alpha_1 + \tfrac{29}{42}(\alpha_2+\alpha_4)\right)\phi\langle \mathbf{e} \rangle,}
\end{eqnarray}
{whose sum yields the $\phi$-terms in equation~\eqref{eq:Shear1LC}. The effect of particle addition for each type of anisotropy introduced in equation~\eqref{eq:TotalLCStress} is as follows:
	(i) $\alpha_3$ (stress–strain anisotropy): the contributions due to $\boldsymbol{\Pi}$ and $\boldsymbol{\tau}^{NN}$ partially cancel, leaving $+\tfrac{1}{2}\alpha_3\,\phi$ in equation~\eqref{eq:Shear1LC}, i.e., a net increase in the $\alpha_3$ contribution in the presence of particles.
	(ii) $\alpha_2$ (bending-resistance anisotropy): since the director is aligned with the flow in the imposed shear, the particle-free LC shows no $\alpha_2$ stress. However, the particle-induced disturbance activates a positive $+\tfrac{15}{7}\alpha_2\,\phi$ via $\boldsymbol{\tau}^{NN}$ that outweighs the negative $-\tfrac{23}{14}\alpha_2\,\phi$ from $\boldsymbol{\Pi}$, giving the net $+\tfrac{1}{2}\alpha_2\,\phi$.
	(iii) $\alpha_1$ (extension along $\mathbf{n}$): weak extensional/compressional lobes in the Newtonian flow disturbance around the sphere also activate a small positive $+\tfrac{25}{126}\alpha_1\,\phi$, absent in the particle-free LC with $\mathbf{n}$ along the flow, through additive contributions from both $\boldsymbol{\Pi}$ and $\boldsymbol{\tau}^{NN}$.
	(iv) $(\alpha_2+\alpha_4)$ (symmetric, director-biased viscosity): strongly sampled by gradients of $u_1$ tangential to the particle, producing the largest enhancement $+\tfrac{25}{12}(\alpha_2+\alpha_4)\,\phi$, with approximately twice the contribution from $\boldsymbol{\Pi}$ as from $\boldsymbol{\tau}^{NN}$ ($\tfrac{39}{28}$ vs.\ $\tfrac{29}{42}$).}

For shear flow with the director aligned along the vorticity direction, the suspension stress becomes:
\begin{equation}\label{eq:Shear3LC}
	\langle\hat{\boldsymbol{\sigma}}\rangle_\text{shear, n$_\text{vorticity}$}=\Big((2+5\phi)+\phi\big(\tfrac{10}{63}\alpha_1+\tfrac{5}{6}(\alpha_2+\alpha_4)\big)\Big)\langle\mathbf{e}\rangle.
\end{equation} 
{The particle-free LC stress is effectively Newtonian here. However, spheres interact with the $\alpha_1$ and $(\alpha_2+\alpha_4)$ anisotropies to activate additional stresses in the suspension. This occurs because the sphere introduces out-of-plane gradients (e.g., $\partial u_1/\partial x_3$) in its disturbance field, creating nonzero projections along the director even though the base shear does not. Accordingly:
	(i) For $\alpha_1$, the entire $+\tfrac{10}{63}\alpha_1\,\phi$ arises from $\boldsymbol{\tau}^{NN}$ (the $\boldsymbol{\Pi}$ part is zero).
	(ii) For $(\alpha_2+\alpha_4)$, the total $+\tfrac{5}{6}(\alpha_2+\alpha_4)\,\phi$ splits into $\tfrac{3}{14}$ from $\boldsymbol{\Pi}$ and $\tfrac{13}{21}$ from $\boldsymbol{\tau}^{NN}$. Thus, the Newtonian disturbance dominates the interaction stresslet in this orientation.}

In summary, the presence of particles in nematic-based DIW inks significantly influences shear and extensional rheology via interactions with the anisotropic microstructure. This interplay can alter flow rates and deposition performance in practical printing applications.

\section{Conclusion}\label{sec:Conclusions}
{In this work, we formalized the ensemble-averaged stress for dilute suspensions in non-Newtonian fluids and introduced a stresslet decomposition that expresses the interaction stresslet as an explicit functional of the non-Newtonian stress alone. Building upon the ensemble-averaging framework of Koch et al.~\cite{koch2016stress} for suspensions in polymeric fluids, our formulation applies broadly to fluids whose total stress admits a Newtonian\,+\,non-Newtonian split. Additionally, in the weakly non-Newtonian regime, where the extra stress is formally $\mathcal{O}(\epsilon)$ relative to the Newtonian contribution, we combine regular perturbation theory with the method of characteristics to evaluate all $\mathcal{O}(\phi\,\epsilon)$ corrections to the suspension stress without solving the first-order momentum equations, yielding substantial computational savings. We demonstrated the approach on two model systems: a spheroidal fluid and a weakly anisotropic nematic liquid crystal, obtaining analytic or semi-analytic expressions and clarifying the physical origin of the particle–microstructure contributions. The method can also be applied to novel fluid systems to enable rapid early-stage screening of materials without resolving first-order flow fields.}

In section~\ref{sec:FormulationAllNonNewtonianFluid}, we derived the general expression for the ensemble-averaged stress in a dilute suspension, based on a decomposition of the fluid stress into Newtonian and non-Newtonian components. This framework is applicable to a wide range of fluids even when the microstructural constituents (e.g., polymers in polymer melts) are not explicitly separable from a Newtonian solvent. The total stress includes contributions from the imposed fluid stress (both Newtonian and non-Newtonian), the particle stresslet (arising from both fluid components), and the particle-induced non-Newtonian stress (PINNS), which captures microstructural distortions caused by particle-induced flow. The combined contribution of the non-Newtonian stresslet and the Newtonian stresslet driven by the non-Newtonian flow is termed the "interaction stresslet." Koch et al.~\cite{koch2016stress} showed that the linearized (about the flow undisturbed by the particles) part of the non-Newtonian stress results in divergent volume integrals. They demonstrated that the ensemble average of the linearized component vanishes and must be subtracted before applying volume averaging. We generalized this remedy to a broader class of non-Newtonian models.

Additionally, we introduced a novel decomposition of the stresslet using the divergence theorem and a generalized reciprocal identity. This separates the interaction stresslet into an “undisturbed” part, corresponding to the stresslet on a particle with undisturbed non-Newtonian stress on its surface, and a volumetric component capturing the deviation of fluid stress from the undisturbed stress in the entire fluid volume. Crucially, the interaction stresslet can be expressed entirely in terms of the non-Newtonian stress field. When the fluid is weakly non-Newtonian, the entire leading-order contribution to the suspension stress can be computed using only the Newtonian (Stokes) velocity field around the particle. This eliminates the need for solving the coupled partial differential equations governing non-Newtonian velocity fields, enabling computationally inexpensive evaluation of effect of particle–microstructure interactions on suspension stress without neglecting any key physical features such as the polymer relaxation in viscoelastic fluids.

We applied this framework to two examples in section~\ref{sec:examples}. The first involved a spheroidal fluid, a dilute suspension of small, non-Brownian spheroids forming the fluid’s microstructure. These spheroids, assumed too small to perturb the flow around the larger suspended spheres, contributed a non-Newtonian stress proportional to their concentration, which served as the perturbation parameter $\epsilon$. We computed the interaction stress as a function of the spheroid aspect ratio $\kappa$, covering thin discs ($\kappa \ll 1$), spheres ($\kappa = 1$), and slender rods ($\kappa \gg 1$). At $\kappa = 1$, the interaction stress arises solely from the stresslet. As $\kappa \to 0$ or $\kappa \to \infty$, we observed a net negative interaction stress dominated by PINNS. This behavior closely parallels the response of suspensions in viscoelastic fluids~\cite{sharma2023steady}, and adds to existing analogies such as the effect of spheroidal fluids on bacterial motility~\cite{kamdar2022colloidal}.

For prolate spheroids ($\kappa > 1$), the negative PINNS arises in regions near the sphere where local orientation deviates from the undisturbed state. These deviations result from misalignment between the local and imposed extensional axes, caused by the dipolar disturbance field generated by the sphere. As $\kappa$ decreases, spheroids reorient more slowly, enlarging the misaligned region. A similar trend is observed in oblate spheroidal fluids with decreasing $\kappa$. These findings show that adding spheres to a spheroidal fluid of aspect ratio greater than about 3.9 or less than 0.36 reduces the overall extensional stress. {These results may also be viewed as providing insight into bidisperse suspensions, particularly regarding the effect of adding smaller spheroidal particles to a dilute suspension of spheres.}

{In the second example, we consider a nematic liquid crystal (LC) with weak mechanical anisotropy. The LC stress, modeled using the Leslie–Ericksen formulation, includes four distinct non-Newtonian contributions: stress–strain anisotropy (the $\alpha_3$ term), bending-resistance anisotropy (the $\alpha_2$ term), elongational response along the director (the $\alpha_1$ term), and a symmetric, director-biased contribution proportional to $\alpha_2+\alpha_4$ (see equation~\eqref{eq:TotalLCStress}). To simplify the analysis, we assume a fixed director field that is not altered by the presence of suspended particles, thereby neglecting particle-induced distortions such as Saturn-ring defects. Under this assumption, the nonlinear component of the non-Newtonian stress vanishes (i.e., $\boldsymbol{\Pi}^{(NL)}=0$, so PINNS $=0$), but the interaction stresslet remains finite and yields a nonzero particle–nematic contribution to the suspension stress. In uniaxial extensional flow (equation~\eqref{eq:ExtensionalLC}), the interaction stresslet associated with the $\alpha_2$ and $\alpha_3$ mechanisms vanishes—even though these terms contribute finite stress in the particle-free case. By contrast, the stresslets proportional to $\alpha_1$ and $\alpha_2+\alpha_4$ enhance the stress, with the same sign as their particle-free counterparts. In simple shear flow, the interaction depends on the director orientation. When the director is aligned with the flow direction (equation~\eqref{eq:Shear1LC}), all four mechanisms contribute to the stress. For the director aligned with the vorticity direction (equation~\eqref{eq:Shear3LC}), the particle–fluid interaction activates stresses due to the $\alpha_1$ and $\alpha_2+\alpha_4$ terms even though these terms do not add to the particle-free stress beyond the Newtonian contribution. The anisotropic microstructure thus selectively modulates the suspension rheology depending on the flow type and director configuration. Future studies that account for director distortions induced by suspended particles will be essential for fully characterizing the rheological response of such systems.}

Together, these examples demonstrate how weakly non-Newtonian microstructures mediate complex particle–fluid interactions in dilute suspensions. The proposed framework enables efficient, physically insightful computation of such effects and can inform the design of advanced materials with tunable rheology, including those used in applications such as direct ink writing, additive manufacturing, and extrusion-based polymer processing.

\section*{Funding}
This work was supported by NSF grant number 2206851.

\bibliographystyle{vancouver}
\bibliography{NonNewtSuspension}
\end{document}